\documentclass[a4paper,11pt]{article}
\usepackage{physics}
\pdfoutput=1 % if your are submitting a pdflatex (i.e. if you have % images in pdf, png or jpg format)

\usepackage{jheppub} % for details on the use of the package, please see the JHEP-author-manual

\usepackage[T1]{fontenc} % if needed
\usepackage{pdfpages}
\usepackage{float}
\usepackage{amsmath,amssymb}
\usepackage[multiple]{footmisc}
\title{\boldmath Trans-Planckian Censorship and the Swampland}
\usepackage{ragged2e}

\author{Alek Bedroya and Cumrun Vafa}

\affiliation{Jefferson Physical Laboratory, Harvard University, Cambridge, MA 02138, USA}

\abstract{In this paper, we propose a new Swampland condition, the Trans-Planckian Censorship Conjecture (TCC), based on the idea that in a consistent quantum theory of gravity sub-Planckian quantum fluctuations should remain quantum and never become larger than the Hubble horizon and freeze in an expanding universe. TCC leads to conditions that are similar to the refined dS Swampland conjecture. For example, applied to the case of cosmologies driven only by a scalar field, the TCC imposes an upper bound of $2/\sqrt{d-2}$ on the asymptotic value of $|V'|/V$. Additionally, it implies that a monotonically decreasing potential across $[\phi_1,\phi_2]$ satisfies $V(\phi_2)\leq A\cdot\exp(-2(\phi_2-\phi_1))/\sqrt{(d-1)(d-2)})$ for some $\mathcal{O}(1)$ constant $A$. Like the dS Swampland conjecture, the TCC forbids long-lived meta-stable dS spaces, but allows sufficiently short-lived ones.}

\begin{document}
\maketitle 
\flushleft
\flushbottom
\setlength{\parindent}{1em}
\setlength{\parskip}{1em}
\section{Introduction}
\justifying
One of the most important challenges facing any fundamental theory of quantum gravity is how to reconcile it with the observed dark energy in our universe. The simplest possibility would be to look for a positive cosmological constant as the background describing our universe. This would necessitate that de Sitter space can exist in such a quantum theory. It has been difficult if not impossible to construct dS spaces (even meta-stable dS) from string theory, which is currently the only well-developed quantum gravitational theory. In fact, the difficulty in constructing dS spaces in string theory was one of the main motivations for the de Sitter Swampland conjecture \cite{obied2018sitter}, which implies that not only there are no meta-stable dS spaces, but also that the slope of the potential satisfies $|V'|\geq c V$ for some constant $c$. In \cite{obied2018sitter}, the idea that $c$ could itself be a function $c(V)$ was contemplated, but all the known string constructions at weak coupling led to the formula with $c$ being a constant independent of $V$, which motivated the dS conjecture.
 There have been refinements of this proposal suggested in \cite{andriot2018sitter, dvali2019exclusion, Andriot:2018mav, garg2018bounds, ooguri2019distance, rudelius2019conditions}.
The conjecture in \cite{obied2018sitter} was mainly due to observations of the structure of the scalar potentials one obtains in string theory, for which with current techniques we have information only at weak couplings which corresponds to large scalar field expectation values. What is missing in the original conjecture is an explanation of it based on fundamental aspects of quantum gravity, such as dynamics of the black holes, as is the case for many other Swampland conjectures. This was partially remedied in \cite{ooguri2019distance} where it was argued why the large field version of the conjecture should hold based on entropy considerations of quasi-dS spaces. Other attempts at coming up with a dS conjecture motivated by more basic aspects of quantum gravity includes \cite{dvali2019exclusion,dvali2019quantum} where quantum breaking of dS is suggested as the main principle leading to $c$ being proportional to $V$ instead of being a constant and \cite{rudelius2019conditions} where the postulate of lack of existence of eternal inflation\footnote{See however \cite{Blanco-Pillado:2019tdf} for a discussion of this.} has led to $c$ being proportional to $V^{1/2}$. However, these are strictly weaker than what one finds in string constructions at 
weak couplings where $c$ is a constant independent of $V$.

It is thus natural to ask if there is any principle of quantum gravity which leads to the dS conjecture at least in large field range but also has specific predictions for any field range. This paper aims to propose such a principle. The principle we propose, the Trans-Planckian Censorship Conjecture (TCC), simply put states that in an expanding universe that could realize in a consistent quantum gravity theory, the sub-Planckian quantum fluctuations should remain quantum and can never become larger than the Hubble horizon and classically freeze\footnote{This notion is different from the similarly named phenomenon discussed in \cite{dolan2017transplanckian,draper2019gravitational}.}. We show that TCC is weaker than the dS Swampland conjecture, but in a way, it is more specific. For example we show that in d-dimensional spacetimes for large field ranges with positive potential $[|V'| \geq c V]\big|_\infty$ with $c_\text{asymptotic}=2/\sqrt{d-2}$. This result is based on the assumption that the cosmological evolution at future infinity is driven only by scalar fields. Additionally, TCC leads to a similar, but weaker, condition in the interior of the field space. We show that TCC requires any monotonically decreasing stretch of the scalar potential $[\phi_i,\phi]$ to obey $V(\phi)\leq A~\exp(-2(\phi-\phi_i))/\sqrt{(d-1)(d-2)})$, where $A$ is an $\mathcal{O}(1)$ constant. This condition tells us that the potential is bounded above by an exponential potential with rate $c_\text{interior}= 2/\sqrt{(d-1)(d-2)}$. This value of $c$ is compatible with all known examples in string theory. However, as we will see, TCC is weaker than dS conjecture at the interior of the field space. In particular, the lower bound for slope $|V'|/V$ depends on the range of the field. Moreover, taking into account quantum fluctuations, TCC is compatible with $V'=0$ points as well, as long as it is sufficiently unstable quantum mechanically. We find that in a meta-stable dS point is compatible with TCC as long as its lifetime $T$ is bounded by
\begin{align}\label{LTI}
T\leq {1\over H} {\rm log} {M_p\over H}
\end{align}
where $H$ is the Hubble parameter and is related to the cosmological constant by ${(d-1)(d-2)\over 2} H^2=V=\Lambda$ in $d$ spacetime dimensions. Also, for unstable critical points, we find a condition similar to the refined dS conjecture which puts a bound on $|V''|/V$ \cite{ooguri2019distance}. Moreover, we find that for any expansionary period of the universe for matter with equation of state $w\geq -1$, measurement of $H$ will give an upper bound to the age of the observed universe. The upper bound is the same as the \eqref{LTI} with $H$ being the measured value of the Hubble parameter at time $T$ after the expansion started.

One of the motivations for this conjecture arises from the issues encountered in the context of studying inflationary models.
One of the most significant triumphs of inflation is the relation that it establishes between quantum fluctuations in the inflationary era and classical macroscopic perturbations that are observable in late-time cosmology. Moreover, it allows for field-theory computations without relying on any trans-Planckian physics. However, it has been pointed out that this framework could fall apart if the mentioned macroscopic fluctuations trace back to trans-Planckian wavelengths during inflation \cite{martin2001trans,brandenberger2001robustness,brandenberger2013trans} (see also \cite{kaloper2003initial,easther2002generic}). In that case, the evolution of fluctuations cannot be reliably extracted from the effective field theory. This issue is called the `trans-Planckian problem' in cosmology literature. Even so, this was not viewed as an obstacle for having such potentials, but only the existence of difficulty in reliably extracting the physics of sub-Planckian fluctuations that cross the horizon from inflationary models for such cases. Here we are proposing that this may never happen and such potentials belong to the Swampland!  

The organization of this paper is as follows: In section 2 we formulate the conjecture and draw some general consequences of it.
In section 3 we study more detailed consequences of this conjecture, in the long field range as well as the short field range. In section 4 we study the consequences of TCC for meta-stable as well as unstable critical points of $V$. In section 5 we present examples from string theory to test TCC.
 In section 6 we discuss possible relations to the distance conjecture. In section 7 we summarize the results and compare with the refined dS conjecture. In section 8 we present our conclusions. Some technical computations are presented in the appendices.

Note Added: In the current version of this article, the interpretation of TCC has been updated compared to the previous versions, without changing any of the equations. In particular, we have highlighted the difference between the implications of TCC for an upper bound on the potential in the interior of the field space and its behavior in the asymptotic field space. This reinterpretation is largely motivated by the results of \cite{vandeHeisteeg:2023uxj} which provides an independent rationale for the result obtained from TCC in the interior of the field space. 

To summarize, in both the interior and the asymptotics of the field space, we find conditions which bear resemblance to that of de Sitter conjecture. However, the corresponding exponential rates are different. For the interior of the field space, we find that a monotonically decreasing positive scalar potential satisfies 
\begin{align}
V(\phi)\leq A~ e^{-c_\text{interior} \Delta\phi},
\end{align}
where $c_\text{interior}=2/\sqrt{(d-1)(d-2)}$, and $A$ is an $\mathcal{O}(1)$ constant while in the asymptotic of the field space, for cosmologies driven by scalar fields, we find
\begin{align}
\frac{|\nabla V|}{V}\geq c_\text{asymptotic},
\end{align}
in Planck units where $c_\text{asymptotic}=2/\sqrt{d-2}$.

\section{The Trans-Planckian Censorship Conjecture (TCC)}
\subsection{Motivations for TCC}

In a quantum gravitational theory, we do not believe that the notion of spacetime as a continuum would make sense at distance scales smaller than Planck length. However, in such a theory we can nevertheless have expansions in the background, which raises the question of what happens to these scales becoming larger than Planck length. In a consistent QG theory, the quantum fluctuations of this kind should remain quantum, in a way not to be contradictory with a classical picture of spacetime at larger scales. However, as is known in the context of inflationary models, when sub-Planckian quantum fluctuations become larger than the Hubble horizon $1/H$, they can become classical and freeze.  This would lead to the classical observation of a sub-Planckian quantum mode, which is a bit strange! This is known as the inflationary trans-Planckian problem \cite{martin2001trans,brandenberger2001robustness,brandenberger2013trans,kaloper2003initial,easther2002generic}.  The traditional view of this problem has been that either we need more information to figure out what happens to these modes or that the structure of the quantum gravitational theory would give the same answer as if the modes were smooth even in the trans-Planckian domain. Here we would like to propose an alternative viewpoint: That such questions should never arise in a consistent quantum gravitational theory! That no trajectory of a consistent quantum theory of gravity should lead to a classical blow-up of the sub-Planckian modes to become larger than the Hubble horizon $1/H$ and that all the QFT's that do lead to this scenario belong to the Swampland.

\subsection{Statement of TCC}

\it We conjecture that a field theory consistent with a quantum theory of gravity does not lead to a cosmological expansion where any perturbation with length scale greater than the Hubble radius trace back to trans-Planckian scales at an earlier time. This could be formulated in terms of initial and final scale factors, $a_i$ and $a_f$ , and final Hubble parameter $H_f$ as
\begin{align}\label{0}
\frac{a_f}{a_i}\cdot l_{pl}<\frac{1}{H_f} \Rightarrow \int_{t_i}^{t_f} Hdt <{\rm \ln} {M_{pl}\over H_f}.
\end{align}
\normalfont
Note that if we take $l_{pl}\rightarrow 0$ or equivalently $M_{pl}\rightarrow \infty$ this condition becomes trivial, as it should with any Swampland condition.
In the following we set (the reduced Planck mass) $M_{pl}\rightarrow 1$.\footnote{Perhaps, a more accurate statement would be to say $\frac{a_f}{a_i}<\frac{KM_{pl}}{H_f}$ for some $\mathcal{O}(1)$ constant $K$. However, unlike other Swampland conjectures which depend on some $\mathcal{O}(1)$ constants, the consequences of TCC are rather insensitive to the exact value of $K$ as it usually appears as a logarithmic correction. Therefore, in this paper, we set $K$ equal to $1$, but one can easily restore the $K$-dependence in all of the results.}\footnote{Under time-reversal, the statement \eqref{0} of TCC for expanding universes, transforms into the following statement for contracting universes.  A field theory consistent with a quantum theory of gravity does not lead to a cosmological contraction where any perturbation with length scale larger than the Hubble scale ($-1/H$) evolve into the sub-Planckian scales at a later time. This could be mathematically formulated in the form in reduced Planck units.
$\frac{a_i}{a_f}<-\frac{1}{H_i}.$}

Since the fluctuations growing bigger than the Hubble radius freeze out, if the wavelength of sub-Planckian quantum fluctuations become larger than the Hubble-radius they turn into classical non-dynamical fluctuations. This leads to the following equivalent statement of TCC in terms of the quantum fluctuations.

\large\bf An equivalent statement of TCC:\normalfont\normalsize

\it Sub-Planckian quantum fluctuations should remain quantum. 
\normalfont

\subsection{Immediate Consequences}

\bf Upper bound on H\normalfont\normalsize

Perhaps, the most immediate consequence of the conjecture \eqref{0} is that for the field theory description to not break down, $H$ must be smaller than $1$ at all times. This is natural as the Hubble parameter is usually proportional to the energy density which must be smaller than Planck energy density for the field theory description to be valid. 

\bf Upper bound on lifetime\normalfont\normalsize

Suppose the equation of state $w=p/\rho$ is greater than $-1$, we can show that the lifetime of universe beginning from $t=t_i$ when it started expanding could be bounded from above by its current value of Hubble parameter, $H_f$. Note that for any combination of conventional matter and radiation, cosmological constant and all of the quintessence models the assumption $w\geq-1$ holds\footnote{This may in principle be violated for phases involving extended objects.}. The rate of change of the Hubble parameter in terms of the energy density $\rho$ and the equation of state $w$ is given by,
\begin{align}\label{Hdec}
\dot H=-(1+w)\frac{\rho}{d-2}.
\end{align}
For $w\geq-1$, the above equation would imply that $H$ is monotonically decreasing. Therefore, for every co-moving time interval $[t_i,t_f]$, we have
\begin{align}
H_fT\leq\int_{t_i}^{t_f}Hdt=\ln(\frac{a_f}{a_i}),
\end{align}
where $T=t_f-t_i$ is the lifetime and $H_f=H(t_f)$. Using the above inequality to bound the LHS of \eqref{0} leads to
\begin{equation}\label{lifetime}
T\leq H_f^{-1}\ln(H_f^{-1}).
\end{equation}
Note that this could also be viewed as an upper bound $H$ in terms of lifetime $T$. The TCC through the inequality \eqref{lifetime} provides a prediction for the current age of the universe. For $H\approx70(km/s)/Mpc$ this upper bound is $\sim 2$ trillion years which is consistent with the age of our universe.

\bf Decelerating expansions are consistent with TCC\normalfont\normalsize

Following, we give a general argument why violating TCC requires accelerating expansion or trans-Planckian energy density $H\geq1$. The inequality \eqref{0} could be written as
\begin{align}
\dot a_f<a_i.
\end{align}
Therefore, violation of TCC requires initial and final points where, 
\begin{equation}\label{TCCv2}
\dot a_f\geq a_i.
\end{equation}
Suppose $H$ is smaller than the Planck scale, we know $\dot a_i/a_i=H<1$. If we use this inequality in \eqref{TCCv2}, we find
\begin{equation}
\dot a_f>\dot a_i.
\end{equation}
Therefore, $\int_{t_i}^{t_f}\ddot a=\dot a_f-\dot a_i$ must be positive and there has been accelerating expansion somewhere along the way. 

\bf TCC vs critical points\normalfont\normalsize

Critical points for a scalar field potentials $V(\Phi)$ are classically forbidden. This is because if we set our initial conditions $\Phi_i$ and $\partial_t\Phi_i$ such that $V'(\Phi_i)=0$ and $\partial_t\Phi=0$, the scalar fields will classically stay at the critical point. This would lead to an accelerating expansion with a constant Hubble parameter which would violate TCC. This argument is of course only true if we ignore quantum effects such as quantum fluctuations and quantum tunneling. Such effects can push the system away from the critical points and potentially save TCC from being violated. We will come back to this point in section 4 and will do a more detailed analysis of the consequences of TCC about critical points by taking quantum effects into account.

\section{Consequences of TCC for Scalar Potentials}
In this section, we find some of the consequences of TCC for scalar fields with a potential $V(\phi)$. We assume $V$ is positive and monotonic. As already noted non-monotonic potentials with critical points are forbidden classically but are allowed when we take into account quantum corrections as we will discuss in the next section. We divide our analysis in this section into three parts. First, we study the consequences of TCC for asymptotic behavior (long field ranges) of the single-field potentials. Next, we generalize some of these results to multi-field models. In the end, we study the short-range predictions of the conjecture for single-field potentials.

\subsection{Long-Range Predictions}\label{PRP}

Using the definition of $H=\frac{\dot a}{a}$, we can rewrite the conjecture \eqref{0} in the form
\begin{align}\label{1}
\int_{\phi_i}^{\phi_f}\frac{H}{\dot\phi}d\phi =\int_{t_i}^{t_f} Hdt<-\ln(H_f).
\end{align}
In $d$ spacetime dimensions, the Friedmann equation takes the form
\begin{align}\label{2}
\frac{(d-1)(d-2)}{2}H^2=\frac{1}{2}\dot\phi^2+V,
\end{align}
and the equation of motion takes the form
\begin{equation}\label{eqmot}
\ddot\phi+(d-1)H\dot\phi+V'=0,
\end{equation}
where $V'$ indicates the derivative of $V$ with respect to $\phi$. Note that we are working in the units where the reduced Planck mass ($M_{pl}=\frac{m_{pl}}{\sqrt{8\pi}}$) is equal to 1. Since $V$ in the equation \eqref{2} is positive, we have
\begin{align}\label{2.5}
\frac{H}{|\dot\phi|}>\frac{1}{\sqrt{(d-1)(d-2)}}.
\end{align}
If we use the above lower bound for the integrand in the equation \eqref{1}, we find
\begin{align}
\frac{|\phi_f-\phi_i|}{\sqrt{(d-1)(d-2)}}<-\ln(H_f),
\end{align}
which can be rearranged in the form
\begin{align}\label{3}
H_f<e^{-\frac{|\phi_f-\phi_i|}{\sqrt{(d-1)(d-2)}}}.
\end{align}
Due to the positivity of the kinetic term in the equation\eqref{2}, $V$ is bounded from above by $(d-1)(d-2)H^2/2$. If we combine this upper bound with the inequality \eqref{3}, we find\footnote{One may conclude that since we can take $\phi_i\rightarrow -\infty$ this would imply that $V$ has to vanish. As we shall discuss one cannot start from arbitrarily negative field value $\phi_i$ to reach arbirary $\phi_f$, which is a necessity for this derivation. 
In other words there is a smallest value of $\phi_i$ one has in the above equation to reach a fixed value of $\phi_f$ including arbitrarily large values of $\phi$.}
\begin{align}\label{4}
V(\phi)<Ae^{-\frac{2}{\sqrt{(d-1)(d-2)}}|\phi-\phi_i|},
\end{align}
where, $A=(d-1)(d-2)/2$ is a constant. For definiteness let us assume $V'<0$. We can use the above inequality to find a lower bound for the average of $-V'/V$ over interval $[\phi_i,\phi_f]$ in the field space.
\begin{align}
\expval{\frac{-V'}{V}}\bigg|_{\phi_i}^{\phi_f}&=\frac{1}{\Delta\phi}\int_{\phi_i}^{\phi_f}\frac{-V'}{V}d\phi=\frac{ln(V_i)-ln(V_f)}{\Delta\phi}.\nonumber
\end{align}
If we combine the upper bound \eqref{4} for $V_f$ with the above identity, we find
\begin{align}\label{5}
\expval{\frac{-V'}{V}}\bigg|_{\phi_i}^{\phi_f}>-\frac{B}{\Delta\phi}+\frac{2}{\sqrt{(d-1)(d-2)}},
\end{align}
where, $B=-\ln(V_i)+\ln(A)$ and $\expval{\frac{-V'}{V}}\bigg|_{\phi_i}^{\phi_f}$ is the average of $\frac{-V'}{V}$ over $[\phi_i,\phi_f]$. 

One may worry about the emergence of light states at large distances in field space expected from the Swampland distance conjecture \cite{Ooguri:2006in}. In particular the interactions between $\phi$ and other fields cannot be ignored in this large field limit and the effective field theory of $\phi$ ignoring the other modes would be invalid in such a limit. However, these modifications do not affect the derivation of the inequalities \eqref{4} and \eqref{5} because all we needed to derive these was $(d-1)(d-2)H^2/2>V$ which is true even if we have additional energy contributions to $H$. Therefore, even for values of $\phi$ where the effective field theory breaks down due to the emergence of a tower of light states, the inequalities \eqref{4} and \eqref{5} are still valid. By taking the limit $\phi_i$ and $\phi_f\rightarrow\infty$ in the eq\eqref{5}, we find
\begin{align}\label{6}
(\frac{|V'|}{V})_{\infty}\geq\frac{2}{\sqrt{(d-1)(d-2)}},
\end{align}
where 
\begin{align}
(\frac{|V'|}{V})_{\infty}:=\liminf_{\phi_i\rightarrow\infty}\liminf_{\phi_f\rightarrow\infty}\expval{\frac{-V'}{V}}\bigg|_{\phi_i}^{\phi_f}.
\end{align}
Thus the inequalities \eqref{4} and \eqref{6} are valid for every value of $\phi$, even when the effective field theory breaks down due to the emergence of a tower of light particles. We now study the family of exponential potentials in more details as they frequently appear in the context of string theory. Let $V\propto e^{-\lambda\phi}$. 
\begin{align}\label{6.5}
\frac{d}{d\phi}(\frac{V}{\dot\phi^2})&=\frac{1}{\dot\phi}\frac{d}{dt}(\frac{V}{\dot\phi^2})\nonumber\\
&=\frac{V'}{\dot\phi^2}-2(\frac{\ddot\phi}{\dot\phi^2})(\frac{V}{\dot\phi^2})\nonumber\\
&=\frac{V'}{\dot\phi^2}(1+2(\frac{V}{\dot\phi^2}))+\frac{2(d-1)H}{\dot\phi}(\frac{V}{\dot\phi^2})\nonumber\\
&=-(\frac{V}{\dot\phi^2})\sqrt{1+2(\frac{V}{\dot\phi^2})}(\lambda\sqrt{1+2(\frac{V}{\dot\phi^2})}-2\sqrt{\frac{d-1}{d-2}}),
\end{align}
where in the third line we used the equation of motion \eqref{eqmot}, and in the fourth line we used the Friedmann equation \eqref{2}. We can rewrite the equation \eqref{6.5} in the form
\begin{align}\label{6.7}
x'=-x\sqrt{1+2x}(\lambda\sqrt{1+2x}-2\sqrt{\frac{d-1}{d-2}}),
\end{align}
where $x:=(V/\dot\phi^2)$ and $x'$ represents the derivative of $x$ with respect to $\phi$. The $x$ is related to the equation of state parameter, $w$, as
\begin{equation}
w=\frac{2}{1+2x}-1.
\end{equation}

If $\lambda>2\sqrt{(d-1)/(d-2)}$, the right hand side of the equation \eqref{6.7} is always negative and $x$ decays exponentially to $0$ as a function of $\phi$. For $\lambda<2\sqrt{(d-1)/(d-2)}$, the right hand side of the \eqref{6.7} has a positive root at $x_c=2(d-1)\lambda^{-2}/(d-2)-1/2$. By checking the signs one can see that $x=x_c$ is an attractor solution and $x$ will converge to $x_c$. Plugging $H$ from the equation\eqref{2} into \eqref{1}, leads to the following form for the trans-Planckian censorship conjecture.
\begin{align}\label{11}
\sqrt{\frac{2x+1}{x(d-1)(d-2)}}V(\phi_f)^\frac{1}{2}=H_f<e^{-\int_{\phi_i}^{\phi_f}\frac{H}{\dot\phi}d\phi}=e^{-\int_{\phi_i}^{\phi_f}\sqrt{\frac{1+2x}{(d-1)(d-2)}}d\phi}.
\end{align}
If we look at the above inequality in the limit $\phi\rightarrow\infty$ where $x$ goes to $x_c=2(d-1)\lambda^{-2}/(d-2)-1/2$, we find
\begin{align}
V(\phi)\leq Ae^{-\frac{4}{(d-2)\lambda}(\phi-\phi_i)},
\end{align}
where $A=x_c(d-1)(d-2)/(2x_c+1)$. Since $V\propto e^{-\lambda\phi}$, also decays exponentially, we have
\begin{equation}
\lambda\geq\frac{4}{(d-1)\lambda}\rightarrow\lambda\geq\frac{2}{\sqrt{d-2}}.
\end{equation}
This inequality could be expressed in terms of $x_c$ and $w$ as
\begin{align}\label{SCexp}
x&<x_{TC}=\frac{d-2}{2}\nonumber\\
w&>w_{TC}=\frac{2}{d-1}-1.
\end{align}
Note that for $\lambda>2/\sqrt{d-2}$, in the attractor solution, $aH/a_i$ goes to zero and because of the fast convergence of the solution to the attractor solution, it is always bounded from above by an $\mathcal{O}(1)$ number. Thus, the conjecture \eqref{0} holds for exponential potentials with decay rate $\lambda>2/\sqrt{d-2}$. 

Following we find the lower bounds for $x_c$ and $w$ in order to have inflation ($\ddot a>0$) and we compare them to $x_{TCC}$ and $w_{TCC}$ in arbitrary dimensions, 
\begin{align}
q=\frac{(d-3)\rho+(d-1)p}{(d-1)(d-2)H^2},
\end{align}
where $q=-\frac{\ddot a a}{\dot a^2}$ is the deceleration parameter, $p=\frac{1}{2}\dot\phi^2-V$ is the pressure and $\rho=\frac{1}{2}\dot\phi^2+V$ is the energy density. For $\ddot a$ to be positive, we must have
\begin{align}
x_c>x_{inf}=\frac{d-2}{2},
\end{align}
which can be expressed in terms of the equation of state as
\begin{align}
w<w_{inf}=\frac{2}{d-1}-1.
\end{align}
These are exactly the same values as \eqref{SCexp}. For exponential potentials, it seems that TCC is equivalent to not having long-field accelerating expansion. This relation is consistent with the general result that we proved in section 1 that violation of TCC necessitates accelerating expansion. 

Note that in the above analysis we ignored the effects of the creation of light states which emerge as the field values roll to infinity. These effects would modify both the Friedmann equation \eqref{2} and the equation of motion \eqref{eqmot}. In this regard \eqref{6} is more robust because it allows for the emergence of a tower of light modes.

\subsection{Generalization to Multi-Field Models}\label{MFG}

In this section we study the applicability of our results to multi-field models where the fields take value in an $n$-dimensional manifold $\mathcal{M}$. Let $\{\phi^j\}_{j=1}^{n}$ be coordinates for a local patch and the metric induced by the kinetic term on $\mathcal{M}$ to take the form $ds^2=G_{ij}d\phi^i\phi^j$ in this coordinate system. For a spatially constant field configuration, the Friedmann equation takes the form
\begin{equation}\label{FEMF}
\frac{(d-1)(d-2)}{2}H^2=\frac{G_{ij}\partial_t\phi^i\partial_t\phi^j}{2}+V(\phi).V_i\Delta\phi
\end{equation}
Let $s$ be the Affine parametrization of the solution path such that
\begin{equation}
G_{ij}\partial_s\phi^i(s)\partial_s\phi^j(s)=1.
\end{equation}
We can rewrite \eqref{FEMF} in terms of $s$ as
\begin{equation}
\frac{(d-1)(d-2)}{2}H^2=\frac{1}{2}(\frac{ds}{dt})^2+V(\phi(s)).
\end{equation}
This is exactly the same as the Friedmann equation in the single field case which we used to derive \eqref{4} with $\phi$ being replaced with $s$. Note that we did not need TCC to hold for all initial conditions to derive \eqref{4}, we only needed TCC to hold for one initial condition. Therefore, the results \eqref{4} holds for the multi-field case as well,
\begin{equation}\label{VMF1}
V(s)<Ae^{-\frac{2}{\sqrt{(d-1)(d-2)}}d^s(\phi_i,\phi_f)},
\end{equation}
where $A=(d-1)(d-2)/2$ and $d^s=\int_{\phi_i}^{\phi_f} ds$ is the canonical length of the solution path from $\phi_i$ to $\phi_f$. Let $d(\phi,\phi_f)$ be the canonical length of the geodesic connecting the two points, then we have $d\leq d^s$. Therefore, we can replace $d^s$ in \eqref{VMF1} with $d$ to get
\begin{equation}\label{potexp}
V(s)<Ae^{-\frac{2}{\sqrt{(d-1)(d-2)}}d(\phi_i,\phi_f)}.
\end{equation}
The above inequality holds for any two points $\phi_i$ and $\phi_f$ that can be connected through a solution to the equations of motion such that the potential remains positive along the path. The derivation of \eqref{6} from \eqref{4} extends without any modifications to the multi-field case and gives
\begin{align}\label{dSMF}
(\frac{|V'|}{V})_\infty>\frac{2}{\sqrt{(d-1)(d-2)}},
\end{align}
where $(\frac{|V'|}{V})_\infty$ is defined as $\liminf_{s_i\rightarrow\infty}\liminf_{s_f\rightarrow\infty}\expval{\frac{-V'(\phi(s))}{V(\phi(s))}}_{[s_i,s_f]}$ where $s$ is the canonical Affine parameter for an arbitrary path with infinite length in $\mathcal{M}$. 

Note that the inequality \eqref{potexp} is only applicable to a pair of points $(\phi_i,\phi_f)$ which are connected by a classical solution. Following, we further explore this relationship between the points in $\mathcal{M}$. 

One can define a causal structure on the moduli space based on which initial conditions can evolve into other ones in an expanding universe. Suppose $x$ and $y$ are two points in the moduli space $\mathcal{M}$, we say $x$ causally precedes $y$, if for some $\dot\phi_i^2<\mathcal{\mathcal{O}(1)}$ the initial field configuration $\phi=x$ can evolve into $\phi=y$. We show this by $x\prec y$. The condition $\dot\phi_i^2<\mathcal{O}(1)$ makes sure that the field theory description does not break. 

Due to the dissipative nature of the Friedmann equations, this causal structure is non-commutative. Generally, to go from a point with a lower potential to a point with a higher potential, we might need a trans-Planckian initial condition $\dot\phi$ to overcome the potential difference in the presence of dissipation. In fact, by assuming our energy density must be sub-Planckian ($H<1$), which is a much weaker assumption than the TCC, we can find an upper bound on the field range that the field $\phi$ can climb up a potential hill.

Suppose $\phi(t)$ is climbing up a positive \bf monotonically increasing \normalfont potential $V$ from $\phi_i$ to $\phi_f$, we find an upper bound on $\Delta\phi=\phi_f-\phi_i$.
\begin{align}
\ddot\phi&=-(d-1)H\dot\phi-V'\nonumber\\
&<-(d-1)H\dot\phi\nonumber\\
&<-\sqrt\frac{2(d-1)V}{d-2}\dot\phi\nonumber\\
&<-\sqrt\frac{2(d-1)V_i}{d-2}\dot\phi.
\end{align}
Integrating the above inequality leads to 
\begin{align}
\Delta\dot\phi+\sqrt\frac{2(d-1)V_i}{d-2}\Delta\phi<0.
\end{align}
Since $\dot\phi_i<\sqrt{(d-1)(d-2)/2}$ (this results from $H<1$), we find
\begin{align}\label{pothill}
\Delta\phi<\frac{d-2}{2}\sqrt\frac{1}{V_i}.
\end{align}
Note that the above upper bound only depends on $V_i$ the value of the potential at the initial point. We can use the full power of TCC to derive another upper bound which also depends on $V_f$ the final value of the potential. From the equation \eqref{potexp}, we know that an initial field value cannot be too far, because other wise the upper bound in \eqref{potexp} would be less than $V_f$. This gives
\begin{align}\label{CSM}
\Delta\phi<\frac{\sqrt{(d-1)(d-2)}}{2}\ln(\frac{A}{V_f}).
\end{align}

In fact, this has the same nature as the inequality \eqref{pothill} since typically going back in the solution requires climbing up a potential hill. This obstruction for extending the solution in the field space only in the past direction happens because of the dissipation in our equations. If two points do not satisfy the inequality \eqref{CSM} for any order of them, they are causally unrelated. This could mean that there is a potential barrier between them that is high enough such that climbing it in the presence of dissipation would need trans-planckian initial conditions. Situations like this can naturally happen for two points in opposite asymptotic regions of the Moduli space, as the potential is highest in the interior and decays exponentially at infinity. 

\begin{figure}
\includegraphics[scale=0.6]{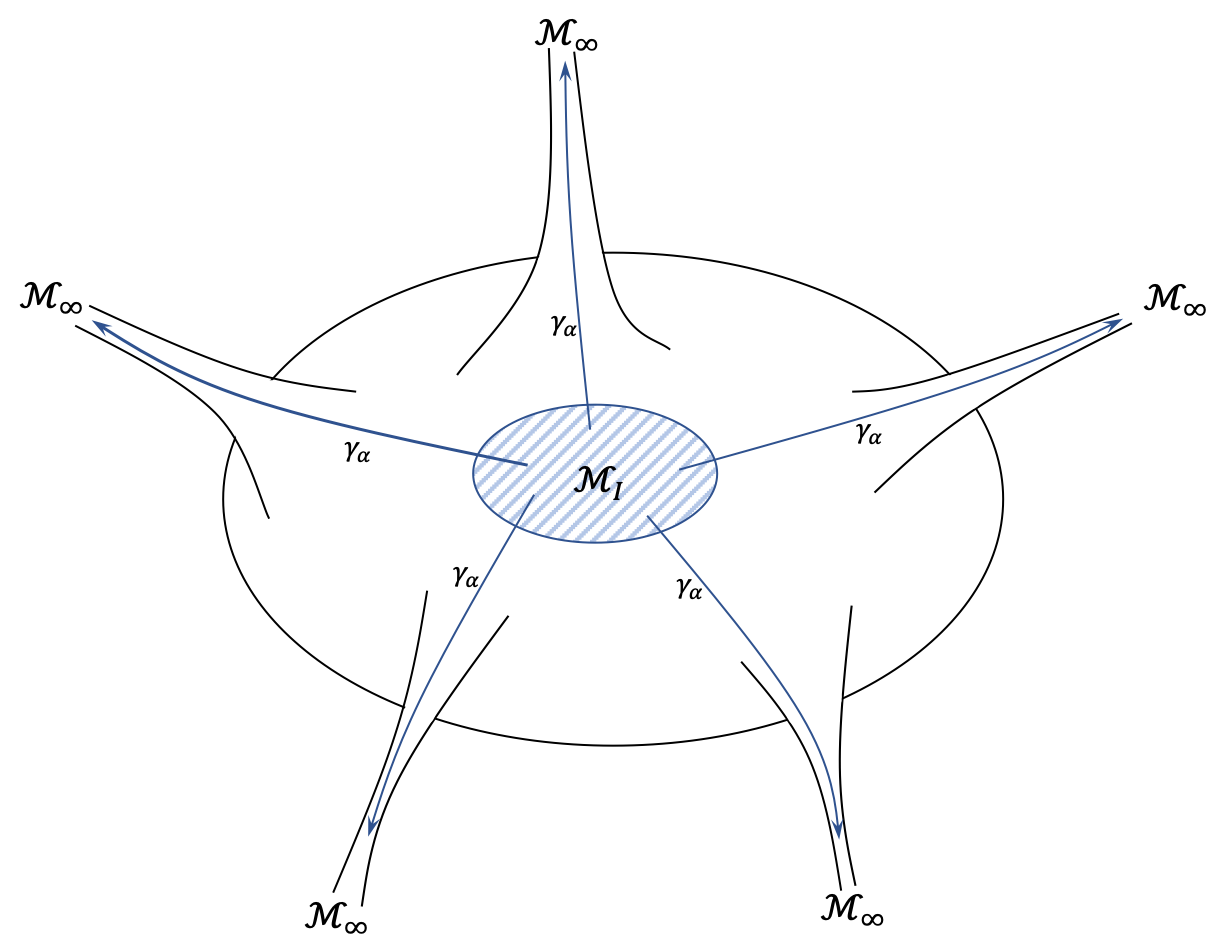}
\centering
\caption{The curves $\gamma_\alpha$ are causal curves that initiate in the interior region $\mathcal{M}_I$ and collectively span the asymptotic region $\mathcal{M}_\infty$.}
\label{Moduli}
\end{figure}

We can use this result to obtain a bound on the asymptotic gradient of the potential.  
We divide the moduli space into two parts, the interior $\mathcal{M}_{I}$ that contains all the local maxima of $V$ and the asymptotic region $\mathcal{M}_{\infty}$ which is located far enough from $\mathcal{M}_I$ with respect to the canonical distance given by the metric defined on $\mathcal{M}$. Since $\mathcal{M}_{I}$ contains the critical points, the causal paths initiated from $\mathcal{M}_I$ can cover all of the moduli space including $\mathcal{M}_\infty$. Suppose $\mathcal{M}_{\infty}$ can be covered by causal paths $\{\gamma_\alpha\}_{\alpha\in\mathcal{I}}$ (with respect to the causal structure defined in \ref{MFG}) such that 
\begin{itemize}
\item they all initiate in $\mathcal{M}_I$.
\item the path $\gamma_\alpha$ is parametrized by the Affine parameter $s_\alpha$.
\end{itemize}
We call every $\alpha\in\mathcal{I}$ an asymptotic direction of the moduli space (fig \ref{Moduli}).
We define
\begin{align}
(\frac{|\nabla_IV|}{V})_\alpha:=\liminf_{s_{\alpha,i}\rightarrow\infty}\liminf_{s_{\alpha,f}\rightarrow\infty}\expval{\frac{|\partial_{s_\alpha}V(\gamma(s_\alpha))|}{V(\gamma(s_\alpha))}}_{[s_{\alpha,i},s_{\alpha,f}]},
\end{align}
where on the right hand side $\expval{\textunderscore{}}_{[s_{\alpha,i},s_{\alpha,f}]}$ is the average over $[s_{\alpha,i},s_{\alpha,f}]$. This roughly represents the ratio $|V'|/V$ along the asymptotic direction ($\partial_{s_\alpha}\gamma(s_\alpha)$) going outward from the interior. We also define
\begin{align}
(\frac{|\nabla V|}{V})_\alpha:=\liminf_{s_{\alpha,i}\rightarrow\infty}\liminf_{s_{\alpha,f}\rightarrow\infty}\expval{\frac{|\nabla V(\gamma(s_\alpha))|}{V(\gamma(s_\alpha))}}_{[s_{\alpha,i},s_{\alpha,f}]},
\end{align}
which roughly represents the limit of $|\nabla V|/V$ as we go to infinity in the asymptotic direction $\alpha$. From the above definitions we know
\begin{align}\label{100}
(\frac{|\nabla V|}{V})_\alpha\geq(\frac{|\nabla_IV|}{V})_\alpha.
\end{align}
On the other hand, from the inequality \eqref{dSMF}, for every $\alpha$ we have
\begin{align}\label{101}
(\frac{|\nabla_IV|}{V})_\alpha\geq \frac{2}{\sqrt{(d-1)(d-2)}}.
\end{align}
Combining \eqref{100} and \eqref{101} leads to 
\begin{align}\label{asmptds}
(\frac{|\nabla V|}{V})_\alpha\geq\frac{2}{\sqrt{(d-1)(d-2)}},
\end{align}
which has the same form as the dS Swampland conjecture \cite{obied2018sitter} but is for the asymptotic region of the moduli space.

\subsection{Short-Range Predictions}

In this section, we prove several inequalities from TCC for the short-field-range behavior of monotonically decreasing positive potentials. 

\paragraph{Obstruction of flatness}~

The trans-Planckian censorship conjecture clearly forbids a flat potential ($V'=0$) as it can lead to accelerated expansion with a fixed Hubble parameter. In our first result in this subsection, we find an inequality which puts an upper bound on the length of the field range over which $|V'|$ is smaller than a constant. Suppose $|V'|_{\max}$ is the maximum of $|V'(\phi)|$ over $\phi\in[\phi_i,\phi_f]$, we have,
\begin{align}
\frac{d\dot\phi^2}{d\phi}=2\ddot\phi\leq2|V'|\leq2|V'|_{\max},
\end{align}
where we used the \eqref{eqmot} for the first inequality. For the initial conditions $\dot\phi=0$ and $\phi=\phi_i$, integrating the above inequality gives
\begin{align}\label{dpine}
\dot\phi(\phi)=\sqrt{\int_{\phi_i}^{\phi}\frac{d\dot\phi^2(\phi')}{d\phi'} d\phi'}\leq\sqrt{2|V'|_{\max}\Delta\phi},
\end{align}
where $\Delta\phi=\phi-\phi_i$. Using the above inequality in the TCC leads to
\begin{align}
\ln(\sqrt\frac{(d-1)(d-2)}{2V(\phi)})&\geq-\ln(H)\nonumber\\
&>\int_{\phi_i}^{\phi_f}\frac{H}{\dot\phi}d\phi\nonumber\\
&\geq\int_{\phi_i}^{\phi_f}\sqrt\frac{1}{(d-1)(d-2)|V'|_{\max}}\sqrt\frac{V(\phi_i)}{\phi-\phi_i}d\phi\nonumber\\
&=\sqrt\frac{V(\phi_i)\Delta\phi}{4(d-1)(d-2)|V'|_{\max}}, 
\end{align}
where in the first and third lines we used $H^2(d-1)(d-2)/2\geq V$, in the second line we used the TCC, and in the third line we used \eqref{dpine}. We can rearrange the above inequality into the form
\begin{align}\label{LF1}
(\frac{|V'|_{\max}}{V_{\max}})>\frac{(\phi_f-\phi)}{4(d-1)(d-2)}\ln(\sqrt\frac{(d-1)(d-2)}{2V(\phi_f)})^{-2}.
\end {align}
We used the monotonicity to replace $V(\phi_i)$ with $V_{\max}$. Note that $V'$ and $V$ are not evaluated at the same point in \eqref{LF1}. However, for regions where the potential is stable ($V''>0$), both $V$ and $V'$ attain their maximum at the same point $\phi= \phi_i$, and the LHS in \eqref{LF1} becomes a local quantity.

The integration in the statement of TCC makes it a global criterion in terms of the potential. In fact, it is very challenging to obtain a local statement about the potential from TCC, which is why the small field range inequalities are weaker than their long-field-range counterpart derived in the previous subsection. We now provide the results of some numerical analysis which supports this observation.
 
 Let $C(\phi_f):=H_f\frac{a_f}{a_i}$. For the conjecture to be true, $C$ must be bounded from above by an $\mathcal{O}(1)$ constant for any physically allowed initial condition (one that $V_i$ and $|\dot\phi_i|$ are both less than 1). The maximum of $C$ over a field range roughly measures the amount of violation of the conjecture. 

Suppose $\lambda$ is the decay rate of an exponential potential, we showed for $\lambda<\lambda_{TC}=\frac{2}{\sqrt{d-2}}$, the conjecture gets violated at infinity. Below, are the results of investigating the consistency of exponential potentials with the conjecture for all field ranges

1) For any value of $\lambda<\lambda_{TC}$, even though the conjecture is violated at infinity, it seems that the conjecture holds for any initial condition over field range $\Delta\phi \sim \mathcal{O}(1)$, which by violation we mean $C>1$. Surprisingly, this is true even for decay rates as small as $\lambda\sim10^{-3}$ that are in contradiction with the conjecture at large field values.

2) For decay rates $\lambda>\lambda_{TC}$ which the conjecture holds at the limit $\phi\rightarrow\infty$, it seems that it also holds for all field values. More specifically, in 4 dimensions, there are no physically allowed initial conditions that would result in a $C>1$ for any $\lambda>\sqrt2+0.01$.

Conclusions:

1) For exponential potentials, it seems that the conjecture is always satisfied for small field values ($\Delta\phi<\mathcal{O}(1)$) and their consistency with the conjecture is determined based on their large-$\phi$ behavior. In other words, the conjecture becomes more non-trivial at large field values. 

2) The conjecture does not restrict the value of $|V'|/V$ over very small field ranges. We can have potentials with arbitrarily small $\lambda$ that satisfy the conjecture for any physically permissible initial conditions over sufficiently small field ranges $\Delta\phi<<\mathcal{O}(1)$. Therefore, this conjecture does not rule out the quintessence models with small decay rates as long as they only last for $\Delta\phi <\mathcal{O}(1)$. In particular, we have checked that the models discussed in \cite{Agrawal:2018own,Agrawal:2019dlm} where $0 < \lambda \leq 0.6$ are compatible with TCC because the field ranges in those models are sufficiently smaller than Planck.

\paragraph{Accelerating roll}~

In this part, using a different assumption, we find an inequality very similar to \eqref{LF1} for small field regime behavior of the potential. Suppose we have a rolling scalar field with positive $\ddot\phi$ over a field range $[\phi_i,\phi_f]$. The equation of motion \eqref{eqmot} implies
\begin{equation}\label{apdd}
(d-1)H\dot\phi<|V'|. 
\end{equation}
This inequality leads to
\begin{align}
\int_{\phi_i}^{\phi_f}\frac{H}{\dot\phi}d\phi&>\int_{\phi_i}^{\phi_f}\frac{(d-1)H^2}{|V'|}d\phi\nonumber\\
&\geq\int_{\phi_i}^{\phi_f}\frac{2}{d-2}\frac{V}{|V'|}d\phi\nonumber\\
&=\frac{2}{d-2}\expval{\frac{V}{|V'|}}\Delta\phi,
\end{align}
where in the first line we used \eqref{apdd} and in the second line we used $H^2\geq\frac{2V}{(d-1)(d-2)}$ from \eqref{2}. Using the above result, in addition to $H\geq\sqrt\frac{2V}{(d-1)(d-2)}$ in \eqref{1}, one can show
\begin{equation}\label{accroll}
\frac{2}{d-2}\expval{\frac{V}{|V'|}}\Delta\phi\leq\ln\sqrt{\frac{(d-1)(d-2)}{2V}}, 
\end{equation}
where $\expval{\frac{V}{|V'|}}$ is the average of $\frac{V}{|V'|}$ over $[\phi_i,\phi_f]$. 

\paragraph{Strongest consequence of TCC for short-field-range behavior of $V$}~

We finish this subsection by discussing an inequality that is proved in the appendix \ref{A1}. For every pair of non-negative numbers $c_1$ and $c_2$ such that $c_2^2(2+c_1^2)<(d-2)/(d-1)$, we find
\begin{equation}\label{SinTCC}
\min(\frac{V(\phi)}{|V'(\phi)|}c_1,c_2)A_1(c_1,c_2,\phi)<\ln(\frac{A_2}{\sqrt{V(\phi+A_3(c_1,c_2,\phi))}}),
\end{equation}
where the identities \eqref{DefABC}, \eqref{defcg}, and \eqref{Deffg}, provide the definitions of functions $A$,$B$ and $C$. In the derivation of \eqref{SinTCC}, we have not weakened the inequalities for obtaining simpler looking result. That comes at the expense of complexity of our final result which makes it hard to physically interpret for an arbitrary potential. If one is interested in the consistency of a specific class of potentials with TCC, by restricting to that class, the inequality might take a much simpler form. In the appendix \ref{A1}, we discuss how this is the case for convex potentials. Moreover, unlike the original conjecture which must be checked for every initial conditions, \eqref{SinTCC} only depends on the potential and could be checked numerically more easily. The \eqref{SinTCC} is derived by estimating the initial condition that is in most tension with the conjecture and looking at the TCC for that initial condition. 

\section{Critical Points of $V$ and Quantum Instabilities}

\subsection{Metastable dS}

We show that the trans-Planckian censorship conjecture implies that the universe cannot get stuck in a local minimum for $V(\phi)$ for an infinite amount of time. We find an upper bound on the lifetime $\tau$ by which every classical local minimum must decay into another state. Therefore, according to the trans-Planckian censorship conjecture, the potential cannot have a positive minimum, or in other words, $\inf V\leq0$. 

For meta-stable dS we have $\Lambda =(d-1)(d-2)H_{\Lambda}^2/2$. Using \eqref{lifetime} we find
\begin{align}\label{decay}
\tau <\frac{1}{H_\Lambda}\ln(\frac{1}{H_\Lambda}),
\end{align}
 In a quantum theory of gravity, even though dS spaces seem to be impossible to attain as a vacuum, it is not implausible that sufficiently short-lived transient quasi-dS like phases could appear, and TCC allows this. The Hubble time of such a background provides a natural time scale and it is reasonable to expect that the lifetime of such an unstable state to be roughly proportional to this characteristic time scale. Indeed, if our universe is stuck in a metastable minimum with $V=\Lambda\approx2.9\times10^{-122}$, the TCC predicts an upper bound of $\tau <2.4$ trillion years on the lifetime of our universe. Thus also in such a case TCC gives an explanation of the coincidence problem: Not only the age of our universe is related to Hubble time, but its lifetime also cannot exceed the Hubble time, up to log corrections.\footnote{There is an interesting similarity between the upper bound on the dS lifetime predicted by TCC and the scrambling time associated to dS space where we use the scrambling time given by \cite{susskind2011addendum}
 $$\tau_{scrambling}\propto {\ln S\over T} ,$$
 where $T$ and $S$ denote temperature and entropy.
 We see that the upper bound for the lifetime of dS space $\tau_{dS}\sim \tau_{scrambling}$ with the substitutions $T_{dS}={H\over 2\pi}$ and $S_{dS}\sim {1/H^2}$.  We thank J. Maldacena for pointing out this connection.}
 
Note that all the above analysis only applies to local minima with positive values of $V$. For example, for a Harmonic potential $V(\phi)=\phi^2$, from numerical analysis we found that the TCC is satisfied over a field range $[-0.9M_{pl},0.9M_{pl}]$. As the field oscillates about the local minimum within this range, the Hubble friction is strong enough that the field does not get stuck in high $V$ for too long. In other words, thanks to the massless graviton, the energy of $\phi$ gets channeled to the gravity sector fast enough that it does not violate the conjecture.

\subsection{Unstable dS}

In this subsection, we show that for a potential with an unstable local maximum, $|V''|$ cannot be small over a large field interval. In other words, over any field interval around the local maximum, there is a lower bound for $|V''|$ so that the quantum fluctuations could push the field away from the extremum point. Otherwise, the field could stay close to the local maximum for a long enough time that leads to a violation of TCC. First, we provide a more heuristic argument to demonstrate what would go wrong with a quadratic potential over a long field range. Afterward, we give a rigorous argument to prove a sharp inequality from TCC. 

Suppose we have a quadratic potential given by
\begin{align}\label{qap}
V(\phi)=\frac{V''(\phi_0)}{2}(\phi-\phi_0)^2+V(\phi_0),
\end{align}
where $V''(\phi_0)<0$. In \cite{rudelius2019conditions}, for the case of $d=4$, it was shown that a gaussian probability distribution centered at $\phi=\phi_0$ solves the Fokker-Planck equation describing the evolution of quantum fluctuations. That result could be easily generalized to the following solution for any dimension $d>2$.
\begin{align}\label{pdis}
Pr[\phi=\phi_c;t]\propto\frac{\exp[-\frac{\phi_c^2}{2\sigma(t)^2}]}{\sigma(t)},
\end{align}
where
\begin{align}
\sigma(t)\propto\frac{H^2\bigg(e^{\frac{2|V''(\phi_0)|t}{(d-1)H}}-1\bigg)^{1/2}}{\sqrt{|V''(\phi_0)|}}.
\end{align}
Note that the expectation value of $H$ remains constant and equal to $\sqrt{2V(\phi_0)/((d-1)(d-2))}$. If the field range over which \eqref{qap} holds is large enough, the above equation would hold for large $t$. As $t$ goes to infinity, $\sigma(t)$ would exponentially grow like $e^{|V''(\phi_0)|t/[(d-1)H]}$. This leads to a lifetime of $(d-1)H/|V''(\phi_0)|$. Comparing this with the upper bound \eqref{decay} gives
\begin{align}
\frac{|V''(\phi_0)|}{V(\phi_0)}\geq\frac{2}{d-2}\ln(\sqrt\frac{(d-1)(d-2)}{2V})^{-1}.
\end{align}
This heuristic argument tells us that either the field range over which the potential is quadratic is bounded from above, or $|V''|/V$ is bounded from below. Following we give a similar, bu rigorous, statement. Suppose $V(\phi)$ is a positive potential such that $V'(\phi_0)=0$ and for every $\phi\in[\phi_0,\Delta\phi]$, we have $V'(\phi)<0$ and $|V''|\leq|V''|_{max}$. If

$$\Delta\phi\geq \frac{B_1(d)B_2(d)^\frac{3}{4}V_{max}^\frac{d-1}{4}V_{min}^\frac{3}{4}\ln(\frac{B_3(d)}{\sqrt{V_{min}}})^\frac{1}{2}}{V_{min}B_2(d)-|V''|_{max}\ln(\frac{B_3(d)}{\sqrt{V_{min}}})^2}$$,
then
\begin{align}\label{unstable}
\frac{|V''|_{max}}{V_{min}}\geq B_2(d)\ln(\frac{B_3(d)}{\sqrt{V_{min}}})^{-2},
\end{align}
where $V_{max}=V(\phi_0)$ and $V_{min}=V(\phi_0+\Delta\phi)$ are respectively the maximum and the minimum of $V$ over $\phi\in[\phi_0,\Delta\phi]$, and $B_1(d)$, $B_2(d)$, and $B_3(d)$ are $\mathcal{O}(1)$ numbers given by
\begin{align}
B_1(d)=&\frac{\Gamma(\frac{d+1}{2})^\frac{1}{2}2^{1+\frac{d}{4}}}{\pi^\frac{d-1}{4}((d-1)(d-2))^\frac{d-1}{4}},\nonumber\\
B_2(d)=&\frac{4}{(d-1)(d-2)},\nonumber\\
B_3(d)=&\sqrt\frac{(d-1)(d-2)}{2}.
\end{align}
These criteria tell us that if $|V''|$ is small enough over a long enough field range, then $|V''|/V$ is bounded from below by a logarithmic function in $V$. This result is very similar to the refined Swampland dS conjecture with a logarithmic correction. For details of the derivation of this result and its application to quadratic potentials see appendix \ref{Unstable}.

\section{Examples from String theory}

\subsection{KKLT and LVS Scenarios}

Even though the KKLT and LVS scenarios have not yet been fully realized in a concrete string model, it would be interesting to check what the consequences of them may be in the context of the TCC. As we shall see below, even though the asymptotic slope of the potentials are in agreement with TCC, the lifetimes of meta-stable dS in these scenarios are incompatible with TCC. Therefore assuming TCC, either these scenarios are not realizable in string theory, or if they are, there should be other decay channels not considered in the literature which would give it a far shorter lifetime.

For the basic KKLT scenario for a highly warped IIB compactification with anti-D3-branes and NS and RR fluxes the potential for the radial modulus looks like \cite{kachru2003sitter}
\begin{equation}\label{KKLT}
V(\phi)\approx\frac{aA}{2}e^{-ae^{\sqrt\frac{2}{3}\phi}-2\sqrt\frac{2}{3}\phi}(\frac{aA}{3}e^{\sqrt\frac{2}{3}\phi-ae^{\sqrt\frac{2}{3}\phi}}+Ae^{-ae^{\sqrt\frac{2}{3}\phi}}+W_0)+De^{-\sqrt{6}\phi},
\end{equation}
where $\phi$ is the canonical radial modulus, $W_0$ is the value of the tree-level superpotential, $a, A$ and $D$ are constants depending on the details of the model. 

At large field values, the potential \eqref{KKLT} is dominated by the last term and hence decays as $e^{-\sqrt{6}\phi}$. This is consistent with \eqref{asmptds} which followed from TCC. 

For values of $D$ within a specific interval, the potential \eqref{KKLT} can result in a metastable dS. In \cite{kachru2003sitter} the lifetime of such a state was found to be of order $\tau \sim exp(c/V)$ for some $\mathcal{O}(1)$ constant $c$. By comparing this result to \eqref{lifetime}, one can see it significantly violates the trans-Planckian censorship conjecture for small values of $V$ (such as the current value of the cosmological constant). Therefore, although the TCC does allow a short-lived metastable dS, it seems that KKLT scenario which allows long-lived meta-stable state is in conflict with it.

Another scenario for obtaining meta-stable dS is the LVS \cite{quevedo2015local}
where the F-term scalar potential takes the form 
\begin{align}\label{LVSP}
V_F\propto(\frac{K^{S\bar S}|D_SW|^2+K^{a\bar b}D_aW\bar D_{\bar b}\bar W}{\mathcal{V}^2})+(\frac{Ae^{-2a\tau}}{\mathcal{V}}-\frac{Be^{-a\tau}W_0}{\mathcal{V}^2}+\frac{C|W_0|^2}{\mathcal{V}^3}),
\end{align}
where $K_{I\bar J}$ is the K\"ahler metric of the internal manifold, $\mathcal{D}$ is the K\"ahler covariant derivative, and $W$ is the superpotential. It was argued that by fine-tuning the coefficients $A$, $B$, and $C$, we can have a scenario in which the above potential has a positive local minimum with the energy of the order of $\Lambda\approx\mathcal{V}^{-3}$ and lifetime of the order of $e^{\frac{1}{\Lambda}}$ \cite{fernando}. This lifetime is similar to the lifetime computed in the KKLT scenario which we studied in the previous subsection and is likewise in contradiction with TCC. 

Note that at large volumes, the potential \eqref{LVSP} decays like $\exp(-3\sqrt\frac{3}{2}\hat\phi)$ where $\hat\phi=\sqrt{2/3}\ln(\mathcal{V})$ is the canonical radial modulus. This decay rate is greater than $\sqrt{2/3}$ and hence is consistent with the inequality \eqref{asmptds} which was a consequence of TCC. 

\subsection{$O(16)\times O(16)$ Heterotic}

For non-supersymmetric Heterotic string theory constructed by twisting the $E_8\times E_8$, in 10 dimensions, there is a cosmological constant in the string frame \cite{alvarez198616}. That constant changes into an exponential potential for the dilaton when we go to the Einstein frame. In that case, the decay rate is $5/\sqrt{2}$ \cite{obied2018sitter}, which is greater than the lower bound $1/(3\sqrt{2})$ provided by the inequality \eqref{asmptds} as a consequence of TCC in 10 dimensions. 

It was shown in \cite{obied2018sitter} that if we compactify this theory down to $d$ dimensions, we have the following lower bound on $|\nabla V|/V$.
\begin{equation}
\frac{|\nabla V|}{V}\geq\min(2\sqrt{\frac{3d-5}{d-2}},\frac{4\sqrt{2}}{\sqrt{(10-d)(d-2)}}).
\end{equation}
For every $d\geq2$, the above lower bound is greater than (or equal to) the lower bound \eqref{asmptds} that follows from the TCC. Therefore, the asymptotic behavior of the potential in these theories is consistent with TCC. 

The cosmological constant in the 10-dimensional theory comes from the one-loop amplitude which is suppressed by a factor of $g_s^2$ compared to the tree-level amplitude. If the correction was suppressed by the factor of $g_s^\alpha$ instead, after going to the Einstein frame we would have gotten an exponential potential with a decay rate of $(9-2\alpha)/\sqrt2$. This could violate TCC for $\alpha>13/3$. Therefore, even though the TCC is formulated in terms of the low energy effective field theory, it is seemingly sensitive to details of the underlying theory of quantum gravity such as the power of $g_s$ in the leading order string theory perturbative corrections. 

\subsection{No-go Theorems in Type II Theories}

It was shown in \cite{hertzberg2007inflationary}, that upon compactifying IIA with $D6$ and $O6$ planes on a Calabi-Yau threefold, in large volume and weak coupling regime, each one of the planes contributes an exponential term in dilaton and volume moduli to the scalar potential. By maximizing the $|V'|/V$ among different directions in the plane of dilaton and radial moduli, it was shown that $|\nabla V|/V$ is greater than $\sqrt{54/13}$.

The above analysis was generalized in \cite{wrase2010classical} to different $Dp$-branes and $Oq$-planes to prove a set of no-go theorems for classical dS vacua. Based on that work, a set of lower bounds for $|\nabla V|/V$ were calculated in \cite{obied2018sitter}. All of the obtained lower bounds are greater than (or equal to) $\sqrt{2/3}$ and therefore are consistent with the lower bound \eqref{asmptds} that follows from the TCC. Interestingly, for the case of O6-branes without any D6-branes, the lower bound obtained in \cite{obied2018sitter} for $|\nabla V|/V$ is exactly equal to $\sqrt{2/3}$. A similar analysis was done in \cite{andriot2019open} for a family of 4 dimensional supergravity solutions with $Op$ and $Dp$ branes discussed in \cite{andriot2019new}, and all the lower bounds obtained for $|\nabla V|/V$ were greater (and in some cases equal to) $\sqrt{2/3}$. The number $\sqrt{2/3}$ for $|\nabla V|/V$ also shows up in the context of studying the relation between the dS swampland conjecture and dS vacua \cite{danielsson2019quantum}\footnote{In \cite{danielsson2019quantum} it was argued that the Bunch-Davies vacuum is problematic and an alternative vacuum was proposed which depends on a UV cutoff $\Lambda<M_{pl}$. It was shown that if such a background gets realized in an inflationary model with potential $V$, the $|V'|/V$ should be related to $\Lambda$ by $|V'|/V\simeq \sqrt{2/3}(\Lambda/M_{pl}).$ Interestingly, for $\Lambda\simeq M_{pl}$, the above identity reproduces the lower bound \eqref{asmptds} obtained from the TCC.}.

\subsection{Energy Conditions}

If we ignore the $\alpha'$ and $g_S$ corrections, in the absence of branes and orientifolds, it was shown in \cite{obied2018sitter} that the strong and null energy conditions lead to the lower bounds of respectively $\lambda_{SEC}=2\sqrt\frac{D-2}{(D-d)(d-2)}$ and $\lambda_{NEC}=2\sqrt\frac{D-d}{(D-2)(d-2)}$ on $|V'|/V$ for a $D$-dimensional theory compactified down to $d$ dimensions. Both of these lower bounds are greater than \eqref{asmptds} and therefore automatically consistent with the long-field-range prediction of TCC.

\section{TCC versus Distance Swampland Conjecture}

It is natural to expect that TCC is related to the distance conjecture \cite{Ooguri:2006in}. In fact, we would imagine that if we have a tower of particles at large field values, the scale of the potential they generate is $V\sim m^d$. From the inequality \eqref{4}, we know that TCC bounds the potential by $V<A \exp(-\frac{2\Delta\phi}{\sqrt{(d-1)(d-2)}})$. This leads to the prediction that

$$m^d<Ae^{-\frac{2\Delta\phi}{\sqrt{(d-1)(d-2)}}}\rightarrow m<Ae^{-\frac{2\Delta\phi}{d\sqrt{(d-1)(d-2)}}}$$
which thus suggests a lower bound ${{2}\over{d\sqrt{(d-1)(d-2)}}}$ on the exponent in the distance conjecture. It would be interesting to check this in various examples.

\section{TCC versus dS Swampland Conjecture}

The results \eqref{LF1} and \eqref{accroll} for positive potentials resemble the dS Swampland conjecture, but they are weaker for small values of potential due to a logarithmic correction. However, the inequality \eqref{asmptds}, which only applies to the asymptotic regions of the moduli space, has the same form as the dS Swampland conjecture. The occurrence of logarithmic corrections seems to be a hallmark of TCC in the interior of the moduli space and it would be very interesting to further study the nature of these corrections.

For potentials with a positive critical point, the inequality \eqref{unstable} is very similar to the refined dS conjecture \cite{ooguri2019distance}. It says that if $|V''|$ is smaller than $|V''|_{\max}$ over a large enough field range, then $|V''|_{max}/V_{min}$ must be bounded from below by a constant up to a logarithmic correction in $V$. It is remarkable that the consequences of the TCC naturally resemble the criteria of the refined dS Swampland conjecture up to logarithmic corrections in the short-field-range and have the same form as the dS Swampland conjecture for the long-field range. Even though the logarithmic corrections make the inequality weaker, lack of unknown $\mathcal{O}(1)$ constants in TCC makes it an analytically powerful Swampland conjecture. 

Given that the consequences of TCC are very similar to that of refined dS conjecture but weaker for shorter field ranges and stronger for longer field ranges, either one, both, or neither could be true. It is natural to ask what are the consequences of both or only one of them being correct. If TCC is not correct, we have nothing to say in this paper! So let us assume TCC is correct. In this case, there are two possibilities: either refined dS conjecture is true or not. If it is true then, we have learned from TCC that the asymptotic value of the slope that appears in the dS conjecture is fixed to be $c_\infty=2/\sqrt{(d-1)(d-2)}$. Of course, this would not imply the dS conjecture value for $c$ has to be this value, since the slope may vary in different regions. Indeed given the cosmological observations of the present cosmology we know that $c< c_\infty=\sqrt{2\over 3}$ \cite{Agrawal:2018own}. If both conjectures are true, the restrictions on $V$ in the interior of the field space are those imposed by refined dS conjecture as TCC is weaker in the interior.

However, the possibility that only TCC is correct is also an interesting possibility: TCC already explains the strongest evidence for the refined dS conjecture, which is the structure of the asymptotic regions of field space (see in particular \cite{ooguri2019distance}). Moreover, it is more specific for the value of $c$ in this region. Since the interior of field space is not easily accessible in weak coupling computations, it is natural to view that TCC is providing a window into this strongly coupled region.  
One other achievement of the dS conjecture is that it gives an explanation of the coincidence problem of why the dark energy is related to Hubble scale \cite{Agrawal:2018own}, whereas this would be lost if long-lived meta-stable dS were possible. However, as we have noted even though TCC allows metastable dS, its lifetime is necessarily bounded by Hubble time up to small log corrections. Therefore, the coincidence problem, whether the present cosmology is quintessence like or dS like, would be perfectly well explained in the TCC setup. Moreover, the $c_\infty$ obtained from TCC is beautifully consistent with all the known constructions in string theory.
So it seems that if only TCC is correct we still maintain all the nice features of dS conjecture and we, in addition, have a possible first principle explanation based on Planckian physics of why a dS type Swampland conjecture may be true.

\section{Conclusions}
The consequences of the trans-Planckian censorship conjecture are consistent with other Swampland conjectures such as the dS conjecture and the distance conjecture. For short-field-ranges, the TCC has weaker consequences than the refined dS conjecture. However, in the limit $\phi\rightarrow\infty$ it provides an explicit lower bound for the slope of log V. Although TCC does not rule out the possibility of a metastable dS space, it provides a robust and natural upper bound on the lifetime of a dS space. In general, the TCC seems to be a highly well-motivated physical criterion which is not very sensitive to the shape of the potential over very small field ranges. The analytically proven consequences of TCC in this work could be readily checked for an arbitrary potential for the purpose of model building. 

It is natural to consider the cosmological implications of TCC for the early universe cosmology and in particular inflationary models.
As shown in \cite{inflation} TCC places strong restrictions on inflationary models, but it does not rule out scenarios with quasi-static or contracting initial phases such as string gas cosmology \cite{Brandenberger:1988aj} or various bouncing scenarios \cite{Finelli:2001sr,Gasperini:1992em,Khoury:2001wf}.

\acknowledgments
We have greatly benefited from discussions with Robert Brandenberger and Marilena Loverde and their explanations of trans-Planckians issues in inflation as well as for an ongoing joint project on TCC \cite{inflation}. We would also like to thank SCGP where this project was initiated at the 2019 SCGP summer workshop. We would also like to thank P. Agrawal, J. Maldacena, J. McNamara and G. Obied for valuable discussions.

The research of CV is supported
in part by the NSF grant PHY-1719924 and by a grant
from the Simons Foundation (602883, CV).

\appendix
\section{A Strong Short-Field-Range Inequality}\label{A1}

In this appendix, we aim to understand what is the strongest short-field-range statement that TCC would imply for an arbitrary monotonically decreasing positive potential. The conjecture must hold for any physically allowed initial condition (one that $\dot\phi_i<\mathcal{O}(1)$ and $V_i<\mathcal{O}(1)$). To deduce a strong inequality from TCC, we focus on an initial condition that seems to challenge \eqref{1} the most. As $\dot\phi$ appears in the denominator of the LHS, a natural guess for the initial conditions with the most tension with the TCC would be small $\dot\phi_i$. From \eqref{Hdec} one can find that $H$ decreases at a rate proportional to $\dot\phi^2$. Thus, small $\dot\phi_i$ could result in an inflationary universe with a slowly-varying Hubble parameter. If $\dot\phi$ does not grow fast enough, the $\frac{a_f}{a_i}$ inflates exponentially leading to a violation of \eqref{0}. With that in mind, we try to obtain an inequality from TCC for small initial field derivative $\dot\phi_i$.

Suppose $\dot\phi_i>0$ is small enough such that $\ddot\phi_i$ given by the \eqref{eqmot} is positive. Let $\phi^*$ be the smallest $\phi>\phi_i$ where $\ddot\phi$ vanishes (later in the appendix we will prove that such a field value exists and we will provide an upper bound for it). Using \eqref{2.5}, we find
\begin{align}
\ddot\phi=-V'-(d-1)H\dot\phi<-V'-\sqrt\frac{d-1}{d-2}\dot\phi^2.
\end{align}
Since $\dot\phi$ is increasing in the interval $[\phi_i,\phi^*]$, we can use the above inequality to find
\begin{align}
\frac{d\dot\phi}{d\phi}=&\frac{\ddot\phi}{\dot\phi}<\frac{\ddot\phi}{\dot\phi_i}\leq\frac{-V'-\sqrt\frac{d-1}{d-2}\dot\phi_i^2}{\dot\phi_i}.
\end{align}
By integrating the above inequality, we find the following upper bound on $\dot\phi$ for every $\phi\in(\phi_i,\phi^*]$.
\begin{align}\label{21}
\dot\phi<\frac{V(\phi_i)-V(\phi)}{\dot\phi_i}-\sqrt\frac{d-1}{d-2}\dot\phi_i(\phi-\phi_i)+\dot\phi_i.
\end{align}
Plugging the above upper bound on $\dot\phi$ into the equation of motion \eqref{eqmot} and using the inequality $H<H_i$, where $H_i$ is the initial Hubble parameter, we find
\begin{align}
\ddot\phi>\ddot\phi_i+(-V'(\phi)+V'(\phi_i))-(d-1)H_i(\frac{V(\phi_i)-V(\phi)}{\dot\phi_i}-\sqrt\frac{d-1}{d-2}\dot\phi_i(\phi-\phi_i)).
\end{align}
By setting $\phi$ to $\phi^*$, at which $\ddot\phi$ vanishes, we find
\begin{equation}\label{pddpr}
(d-1)H_i(\frac{V(\phi_i)-V(\phi^*)}{\dot\phi_i}-\sqrt\frac{d-1}{d-2}\dot\phi_i\Delta\phi)-(-V'(\phi^*)+V'(\phi_i))>\ddot\phi_i,
\end{equation}
where $\Delta\phi=\phi^*-\phi_i$. According to the mean value theorem, there is a point $\phi_1\in[\phi,\phi^*]$ such that
\begin{align}
(d-1)H_i(\frac{V(\phi_i)-V(\phi)}{\dot\phi_i})-(V'(\phi_i)-V'(\phi^*))=\Delta\phi[(d-1)H_i(\frac{-V'(\phi_1)}{\dot\phi_i})+V''(\phi_1)].
\end{align}
We can rewrite the inequality \eqref{pddpr} in terms of the values of $V'$ and $V''$ at $\phi_1$ as
\begin{align}\label{A7}
\Delta\phi>\frac{-(d-1)H_i\dot\phi_i+|V'(\phi_i)|}{(d-1)H_i(\frac{|V'(\phi_1)|}{\dot\phi_i}-\sqrt\frac{d-1}{d-2}\dot\phi_i)+V''(\phi_1)},
\end{align}
where we used the equation of motion \eqref{eqmot} to substitute $\ddot\phi_i$ for the numerator of the right hand side. Suppose $\dot\phi$ is small enough such that 
\begin{align}\label{criteria1}
\dot\phi_i\leq c_1\sqrt{V(\phi_i)}~\&~\dot\phi\leq c_2\frac{|V'(\phi_i)|}{\sqrt{V(\phi_i)}}~\&~ \dot\phi_i\frac{V''(\phi_1)}{|V'(\phi_1)|}\leq c_3\sqrt{V(\phi_i)},
\end{align}
for some non-negative numbers $c_1$, $c_2$, and $c_3$ satisfying $c_2^2(2+c_1^2)<(d-2)/(d-1)$, we have
\begin{align}\label{c1a}
\dot\phi_i\leq c_1\sqrt{V(\phi_i)}\rightarrow H_i\leq \sqrt\frac{2+c_1^2}{(d-1)(d-2)}\sqrt{V(\phi_i)},
\end{align}
\begin{align}\label{c2a}
\dot\phi_i\leq c_2\frac{|V'(\phi_i)|}{\sqrt{V(\phi_i)}}\rightarrow -(d-1)H_i\dot\phi_i+|V'(\phi_i)|\geq|V'(\phi_i)|(1-c_2\sqrt\frac{(d-1)(2+c_1^2)}{d-2}),
\end{align}
\begin{align}\label{c3a}
\dot\phi_iV''(\phi_1)\leq c_3\sqrt{V(\phi_i)}|V'(\phi_1)|\rightarrow V''(\phi_1)\leq c_3\frac{\sqrt{V(\phi_i)}|V'(\phi_1)|}{\dot\phi_i},
\end{align}
where we used the Friedmann equation\eqref{2} in derivation of \eqref{c1a}, and we used \eqref{c1a} in the derivation of the \eqref{c2a}. Since $\dot\phi_i>0$ we have
\begin{align}\label{intm}
(d-1)H_i(\frac{|V'(\phi_1)|}{\dot\phi_i}-\sqrt\frac{d-1}{d-2}\dot\phi_i)+V''(\phi_1)<(d-1)H_i\frac{|V'(\phi_1)|}{\dot\phi_i}+V''(\phi_1).
\end{align}
By multiplying \eqref{c1a} by $(d-1)|V'(\phi_1)|/\dot\phi_i$ and summing it up with \eqref{c3a} we find that 
\begin{align}
(d-1)H_i\frac{|V'(\phi_1)|}{\dot\phi_i}+V''(\phi_1)\leq(c_3+\sqrt\frac{(d-1)(2+c_1^2)}{d-2})\frac{\sqrt{V(\phi_i)}|V'(\phi_1)|}{\dot\phi_i}.
\end{align}
If we combine this with \eqref{intm}, we find
\begin{align}
(d-1)H_i(\frac{|V'(\phi_1)|}{\dot\phi_i}-\sqrt\frac{d-1}{d-2}\dot\phi_i)+V''(\phi_1)<(c_3+\sqrt\frac{(d-1)(2+c_1^2)}{d-2})\frac{\sqrt{V(\phi_i)}|V'(\phi_1)|}{\dot\phi_i}.
\end{align}
Dividing \eqref{c2a} by the above inequality leads to 
\begin{align}\label{intm2}
\frac{-(d-1)H_i\dot\phi_i+|V'(\phi_i)|}{(d-1)H_i(\frac{|V'(\phi_1)|}{\dot\phi_i}-\sqrt\frac{d-1}{d-2}\dot\phi_i)+V''(\phi_1)}&>\frac{1-c_2\sqrt\frac{(d-1)(2+c_1^2)}{d-2}}{c_3+\sqrt\frac{(d-1)(2+c_1^2)}{d-2}}(\frac{|V'(\phi_i)|}{|V'(\phi_1)|})\frac{\dot\phi_i}{\sqrt{V(\phi_i)}}\nonumber\\
&\geq\frac{1-c_2\sqrt\frac{(d-1)(2+c_1^2)}{d-2}}{c_3+\sqrt\frac{(d-1)(2+c_1^2)}{d-2}}(\frac{|V'(\phi_i)|}{\max_{\phi\in[\phi_i,\phi^*]}(|V'(\phi)|)})\frac{\dot\phi_i}{\sqrt{V(\phi_i)}}\nonumber\\
&=c_2f(c_1,c_2,c_3)g(\dot\phi_i)\frac{\dot\phi_i}{\sqrt{V(\phi_i)}},
\end{align}
where $f(c_1,c_2,c_3)$ and $g(\dot\phi_i)$ are given by
\begin{align}\label{Deffg}
f(c_1,c_2,c_3)&=\frac{1-c_2\sqrt\frac{(d-1)(2+c_1^2)}{d-2}}{c_2(c_3+\sqrt\frac{(d-1)(2+c_1^2)}{d-2})}\nonumber\\
g(\dot\phi_i)&=\frac{|V'(\phi_i)|}{\max_{\phi\in[\phi_i,\phi^*]}(|V'(\phi)|)}.
\end{align}
Using the assumption $\dot\phi\leq c_2|V'(\phi_i)|/\sqrt{V(\phi_i)}$ we can lower the right hand side of \eqref{intm2} to get
\begin{align}
\frac{-(d-1)H_i\dot\phi_i+|V'(\phi_i)|}{(d-1)H_i(\frac{|V'(\phi_1)|}{\dot\phi_i}-\sqrt\frac{d-1}{d-2}\dot\phi_i)+V''(\phi_1)}>f(c_1,c_2,c_3)g(\dot\phi_i)\frac{\dot\phi_i^2}{|V'(\phi_i)|}.
\end{align}
By combining the above inequality with \eqref{A7} we find
\begin{align}\label{Deltapin}
\Delta\phi>f(c_1,c_2,c_3)g(\dot\phi_i)\frac{\dot\phi_i^2}{|V'(\phi_i)|}.
\end{align}
For every $\phi\in[\phi_i,\phi_i+\frac{f(c_1,c_2,c_3)g(\dot\phi)\dot\phi_i^2}{|V'(\dot\phi_i)|}]$ we have
\begin{align}\label{Bin1}
\frac{|V(\phi)-V(\phi_i)|}{\dot\phi_i}&\leq\frac{\phi-\phi_i}{\dot\phi_i}\max_{\phi\in[\phi_i,\phi_i+\frac{f(c_1,c_2,c_3)g(\dot\phi)\dot\phi_i^2}{|V'(\dot\phi_i)|}]}(|V'(\phi)|)\nonumber\\
&\leq\frac{\phi-\phi_i}{\dot\phi_i}\max_{\phi\in[\phi_i,\phi^*]}(|V'(\phi)|)\nonumber\\
&\leq f(c_1,c_2,c_3)g(\dot\phi_i)\frac{\dot\phi_i}{|V'(\dot\phi_i)|}\max_{\phi\in[\phi_i,\phi^*]}(|V'(\phi)|)\nonumber\\
&=f(c_1,c_2,c_3)\dot\phi_i,
\end{align}
where in the first line we used the mean value theorem, in the second line we used \eqref{Deltapin}, and in the third line we used the \eqref{Deffg}, the definition of $g(\dot\phi_i)$. Using the inequalities we have derived, we find
\begin{align}\label{22}
&\frac{\dot\phi_i\sqrt{2}f(c_1,c_2,c_3)g(\dot\phi_i)}{|V'(\phi_i)|\sqrt{(d-1)(d-2)}(f(c_1,c_2,c_3)+1)}\nonumber\\
&~~~~~~~~~~~~~~~~~~~~~~=\int_{\phi_i}^{\phi_i+f(c_1,c_2,c_3)g(\dot\phi_i)\frac{\dot\phi_i^2}{|V'(\phi_i)|}}\frac{\sqrt\frac{2}{(d-1)(d-2)}}{(f(c_1,c_2,c_3)+1)\dot\phi_i}d\phi\nonumber\\
&~~~~~~~~~~~~~~~~~~~~~~\leq \int_{\phi_i}^{\phi_i+f(c_1,c_2,c_3)g(\dot\phi_i)\frac{\dot\phi_i^2}{|V'(\phi_i)|}}\frac{\sqrt\frac{2}{(d-1)(d-2)}}{\frac{|V(\phi)-V(\phi_i)|}{\dot\phi_i}-\sqrt\frac{d-1}{d-2}\dot\phi_i(\phi-\phi_i)+\dot\phi_i}d\phi\nonumber\\
&~~~~~~~~~~~~~~~~~~~~~~\leq \int_{\phi_i}^{\phi_i+f(c_1,c_2,c_3)g(\dot\phi_i)\frac{\dot\phi_i^2}{|V'(\phi_i)|}}\frac{\sqrt\frac{2}{(d-1)(d-2)}}{\dot\phi}d\phi\nonumber\\
&~~~~~~~~~~~~~~~~~~~~~~= \int_{\phi_i}^{\phi_i+f(c_1,c_2,c_3)g(\dot\phi_i)\frac{\dot\phi_i^2}{|V'(\phi_i)|}}\frac{1}{\sqrt{V(\phi)}}\frac{\sqrt\frac{2V(\phi)}{(d-1)(d-2)}}{\dot\phi}d\phi\nonumber\\
&~~~~~~~~~~~~~~~~~~~~~~\leq \frac{1}{\sqrt{V(\phi_i+f(c_1,c_2,c_3)g(\dot\phi_i)\frac{\dot\phi_i^2}{|V'(\phi_i)|})}} \int_{\phi_i}^{\phi_i+f(c_1,c_2,c_3)g(\dot\phi_i)\frac{\dot\phi_i^2}{|V'(\phi_i)|}}\frac{\sqrt\frac{2V(\phi)}{(d-1)(d-2)}}{\dot\phi}d\phi\nonumber\\
&~~~~~~~~~~~~~~~~~~~~~~\leq \frac{1}{\sqrt{V(\phi_i+f(c_1,c_2,c_3)g(\dot\phi_i)\frac{\dot\phi_i^2}{|V'(\phi_i)|})}}\int_{\phi_i}^{\phi_i+f(c_1,c_2,c_3)g(\dot\phi_i)\frac{\dot\phi_i^2}{|V'(\phi_i)|}}\frac{H}{\dot\phi}d\phi\nonumber\\
&~~~~~~~~~~~~~~~~~~~~~~<\frac{1}{\sqrt{V(\phi_i+f(c_1,c_2,c_3)g(\dot\phi_i)\frac{\dot\phi_i^2}{|V'(\phi_i)|})}}\ln(\frac{1}{H(\phi+f(c_1,c_2,c_3)g(\dot\phi_i)\frac{\dot\phi_i^2}{|V'(\phi_i)|})})\nonumber\\
&~~~~~~~~~~~~~~~~~~~~~~\leq \frac{1}{\sqrt{V(\phi_i+f(c_1,c_2,c_3)g(\dot\phi_i)\frac{\dot\phi_i^2}{|V'(\phi_i)|})}}\ln(\frac{\sqrt\frac{(d-1)(d-2)}{2}}{\sqrt{V(\phi+f(c_1,c_2,c_3)g(\dot\phi_i)\frac{\dot\phi_i^2}{|V'(\phi_i)|})}}),
\end{align}
where in the third line we used \eqref{Bin1}, in the fourth line we used \eqref{21}, in the sixth line we used the monotonicity of $V$, in the seventh and the ninth lines we used $V\leq H^2(d-1)(d-2)/2$, and in the eighth line we used the TCC. Below we list the assumptions we made to derive the inequality \eqref{22}. 
\begin{align}\label{criteria2}
&\dot\phi_i\leq \min(c_1\sqrt{V(\phi_i)}, c_2\frac{|V'(\phi_i)|}{\sqrt{V(\phi_i)}}),\nonumber\\
&and\nonumber\\ 
&\dot\phi_i \max_{\phi\in[\phi_i,\phi^*]}(\frac{V''(\phi)}{|V'(\phi)|})\leq c_3\sqrt{V(\phi_i)}.
\end{align}
Following we find an upper bound for $\phi^*$ in terms of $\phi_i$, $\dot\phi_i$ and $V(\phi_i)$ so that by replacing $\phi^*$ in the criteria \eqref{criteria2} we change them into criteria that only depend on the initial conditions. 
\begin{align}
H_i&>H_i-H(\phi^*)\nonumber\\
&=-\int_{\phi_i}^{\phi^*}\frac{\dot H}{\dot\phi}d\phi\nonumber\\
&=\int_{\phi_i}^{\phi^*}\frac{\dot\phi}{d-2}d\phi\nonumber\\
&\geq\frac{\dot\phi_i}{d-2}(\phi^*-\phi_i),
\end{align}
which can be rearranged into the form
\begin{align}\label{UBPS}
\phi^*<\frac{(d-2)H_i}{\dot\phi_i}+\phi_i.
\end{align}
By replacing $\phi^*$ in \eqref{criteria2} with this upper-bound, our criteria change into
\begin{align}\label{criteria}
&\dot\phi_i\leq\min(c_1\sqrt{V(\phi_i)}, c_2\frac{|V'(\phi_i)|}{\sqrt{V(\phi_i)}}),\nonumber\\
&and\nonumber\\ 
&\dot\phi_i \max_{\phi\in[\phi_i,\frac{(d-2)H_i}{\dot\phi_i}+\phi_i]}(\frac{V''(\phi)}{|V'(\phi)|})\leq c_3\sqrt{V(\phi_i)}.
\end{align}
We can view the last inequality as an inequality for $c_3$ rather than a criterion for $\dot\phi_i$. Moreover, it seems that to get the most non-trivial result from the inequality \eqref{22}, we should pick the largest $\dot\phi$ possible. We can choose $\dot\phi$ such that $\dot\phi_i=\min(c_1\sqrt{V(\phi_i)}, c_2\frac{|V'(\phi_i)|}{\sqrt{V(\phi_i)}})$ and then we can pick $c_3$ accordingly as follows to make sure that all of the criteria are satisfied. 
\begin{align}\label{defcg}
\dot\phi_i&=\min(c_1\sqrt{V(\phi_i)}, c_2\frac{|V'(\phi_i)|}{\sqrt{V(\phi_i)}}),\nonumber\\
c_3&=\max(0,\dot\phi_i\max_{\phi\in[\phi_i,\frac{(d-2)H_i}{\dot\phi_i}+\phi_i]}(\frac{V''(\phi)}{|V'(\phi)|})).
\end{align}
From this point on, we take the above identities as definitions of $\dot\phi_i$ and $c_3$. Note that for a given potential $V(\phi)$, $c_3$, and $\dot\phi_i$ are now functions of $\phi_i$, $c_1$ and $c_2$. Therefore from now on, we show them as $c_3(c_1,c_2,\phi_i)$ and $\dot\phi(c_1,c_2,\phi_i)$. By plugging \eqref{defcg} into the inequality \eqref{22}, we find the following two-parameter family of inequalities for non-negative pair of numbers $(c_1,c_2)$ where $c_2^2(2+c_1^2)<(d-2)/(d-1)$. For every $\phi$ we have
\begin{align}\label{Sine2}
\min(\frac{V(\phi)}{|V'(\phi)|}c_1,c_2)A_1(c_1,c_2,\phi)<\sqrt\frac{V(\phi)}{V(\phi+A_3(c_1,c_2,\phi))}\ln(\frac{A_2}{\sqrt{V(\phi+A_3(c_1,c_2,\phi))}}),
\end{align}
where,
\begin{align}\label{DefABC}
A_1&=\frac{f(c_1,c_2,c_3(c_1,c_2,\phi))g(\dot\phi(c_1,c_2,\phi))\sqrt{2}}{\sqrt{(d-1)(d-2)}(1+f(c_1,c_2,c_3(c_1,c_2,\phi)))},\nonumber\\
A_2&=\sqrt\frac{(d-1)(d-2)}{2},\nonumber\\
A_3&=f(c_1,c_2,c_3(c_1,c_2,\phi))g(\dot\phi_i(c_1,c_2,\phi))\frac{\min(c_1\sqrt{V(\phi)}, c_2\frac{|V'(\phi)|}{\sqrt{V(\phi)}})^2}{|V'(\phi)|^2}.
\end{align}
The inequality \eqref{Sine2}, although complicated, is very strong. It is almost local in the sense that it mostly depends on the values of $V$ and its derivatives at point $\phi$, and provides a good way to see if an arbitrary potential violates TCC. This inequality does not depend on initial conditions since we used TCC for the initial conditions that seem to challenge TCC the most to find it. This feature makes it easy to be applied to an arbitrary potential numerically or a class of potentials analytically. For example, for convex potentials \eqref{Sine2} takes much simpler form since $g(c_1,c_2,\phi)=1$. Note that in the case which $c_2$ is large enough such that $V/|V'|$ comes out of the $\min$ function on the LHS of \eqref{Sine2}, we get an inequality very similar to the dS conjecture except an extra logarithmic term. In fact, most of the local results that we find from TCC share this feature. 

\section{Unstable Critical Points}\label{Unstable}

In this appendix we prove the inequality \eqref{unstable} which can be stated as in the following form.

Suppose $\phi_0$ is a critical point (local maximum) of $V(\phi)$, such that $V'<0$ and $|V''(\phi)|\leq |V''|_{\max}$ over the field range $\phi_0\leq\phi\leq\phi_0+\Delta\phi$. Then, either 
\begin{align}\label{Aunst}
\Delta\phi< \frac{B_1(d)B_2(d)^\frac{3}{4}V_{max}^\frac{d-1}{4}V_{min}^\frac{3}{4}\ln(\frac{B_3(d)}{\sqrt{V_{min}}})^\frac{1}{2}}{V_{min}B_2(d)-|V''|_{max}\ln(\frac{B_3(d)}{\sqrt{V_{min}}})^2},~~or~~ \frac{|V''|_{max}}{V_{min}}\geq B_2(d)\ln(\frac{B_3(d)}{\sqrt{V_{min}}})^{-2},
\end{align}
where $V_{max}=V(\phi_0)$ and $V_{min}=V(\phi_0+\Delta\phi)$ are respectively the maximum and the minimum of $V$ over $\phi\in[\phi_0,\Delta\phi]$, and $B_1(d)$, $B_2(d)$, and $B_3(d)$ are $\mathcal{O}(1)$ numbers given by
\begin{align}
B_1(d)=&\frac{\Gamma(\frac{d+1}{2})^\frac{1}{2}2^{1+\frac{d}{4}}}{\pi^\frac{d-1}{4}((d-1)(d-2))^\frac{d-1}{4}},\nonumber\\
B_2(d)=&\frac{4}{(d-1)(d-2)},\nonumber\\
B_3(d)=&\sqrt\frac{(d-1)(d-2)}{2}.
\end{align}

 To show the above result, we prove the following one parameter family of inequalities for $0 \leq c \leq 1$.
\begin{align}\label{Uns2}
&\Delta\phi<\frac{c^{1/2}}{1-c^2}B_1(d)\frac{V_{max}^\frac{d-1}{4}}{|V''|_{max}^\frac{1}{4}}~~~or~~~\frac{|V''|_{max}}{V_{min}}\geq c^2B_2(d)\ln(\frac{B_3(d)}{\sqrt V_{min}})^{-2}.
\end{align}
One can check that by setting $c$ equal to $\min(1,\epsilon+\sqrt\frac{|V''|_{max}\ln(\frac{B_3(d)}{\sqrt{V_{min}}})^2}{V_{min}B_2(d)})$ and taking the limit $\epsilon\rightarrow 0^+$, we can recover the statement \eqref{Aunst}.

\it Proof of \eqref{Uns2}:\normalfont

We start by assuming that first inequality in the \eqref{Uns2} is violated, and will prove that for TCC to hold, the second inequality must be true. Violation of the first inequality implies
\begin{align}\label{assum}
\Delta\phi\geq\frac{c^{1/2}}{1-c^2}B_1(d)\frac{V_{max}^\frac{d-1}{4}}{|V''|_{max}^\frac{1}{4}}.
\end{align}
We treat the problem semi-classically in the sense that we demand the TCC to hold for all classical evolutions with initial conditions
\begin{align}
\phi(t=0)=\phi_0+\delta\phi_i~~~\&~~~\dot\phi(t=0)=\delta\dot\phi_i,
\end{align}
where $\delta\phi_i=\sqrt{\expval{(\phi-\phi_0)^2}}$ and $\delta\dot\phi_i=\sqrt{\expval{\dot\phi^2}}$. In the appendix \ref{AB} we study the quantum fluctuations to find the lower bound on the product $\delta\phi_i\delta\dot\phi_i$. Later, we will optimize our choice of initial conditions among all those that satisfy that uncertainty principle. Until then, we express all of our results in terms of arbitrary initial conditions $\delta\phi_i$ and $\delta\dot\phi_i$. 

From the equation of motion \eqref{eqmot}, we have
\begin{align}
&\ddot\phi\leq\ddot\phi+(d-1)H\dot\phi=-V'\leq|V''|_{max}(\phi-\phi_0),
\end{align}
where in the last inequality we used the mean value theorem. If we use the mean value theorem again, we find
\begin{align}
\ddot\phi\leq|V''|_{max}(\phi-\phi_0)\leq|V''|_{max}t\dot\phi_{max}+\delta\phi_i|V''|_{max},
\end{align}
where $\dot\phi_{max}(t)=max_{t'\in[0,t]}\{\dot\phi\}$. If we integrate this inequality from $t'=0$ to $t'=t$, using $\dot\phi_{max}(t')\leq\dot\phi_{max}(t)$ we find
\begin{align}
\dot\phi\leq\frac{|V''|_{max}}{2}t^2\dot\phi_{max}+|V''|_{max}\delta\phi_it+\delta\dot\phi_i.
\end{align}
Since the right hand side is monotonic in $t$, and the left hand side is equal to $\dot\phi_{max}$ for some $t'\in[0,t]$, we have
\begin{align}
\dot\phi_{max}\leq\frac{|V''|_{max}}{2}t^2\dot\phi_{max}+|V''|_{max}\delta\phi_it+\delta\dot\phi_i.
\end{align}
Suppose $c$ is a positive number smaller than $1$, for $t\leq\sqrt{2/|V''|_{max}}c$, the above inequality gives us
\begin{align}
&\dot\phi_{max}\leq\frac{|V''|_{max}\delta\phi_it+\delta\dot\phi_i}{1-\frac{|V''|_{max}t^2}{2}}\leq\frac{|V''|_{max}\delta\phi_it+\delta\dot\phi_i}{1-c^2}.
\end{align}
From $\dot\phi\leq\dot\phi_{max}$ we find
\begin{align}
\dot\phi\leq\frac{|V''|_{max}\delta\phi_it+\delta\dot\phi_i}{1-c^2}.
\end{align}
Integrating this inequality gives
\begin{align}
\phi-\phi_0\leq\frac{|V''|_{max}t^2\delta\phi_i}{2(1-c^2)}+\frac{\delta\dot\phi_it}{1-c^2}+\delta\phi_i.
\end{align}
Using $t\leq c\sqrt{2/|V''|_{max}}$ again, we find
\begin{align}\label{40}
\phi-\phi_0&\leq(1+\frac{c^2}{1-c^2})\delta\phi_i+\frac{\delta\dot\phi_i}{1-c^2}\frac{c\sqrt2}{\sqrt {|V''|_{max}}}\nonumber\\
&=\frac{2}{1-c^2}\delta\phi_i+\delta\dot\phi_i\frac{c\sqrt2}{(1-c^2)\sqrt{|V''|_{max}}}.
\end{align}
The above inequality is true for all $t\leq c\sqrt{2/|V''|_{max}}$ such that $\phi(t)\leq \phi_0+\Delta\phi$. If the right hand side in \eqref{40} is less than $\Delta\phi$, that would mean $\phi$ is in $[\phi_0,\phi+\Delta\phi]$ for every $t\leq c\sqrt{2/|V''|_{max}}$. We show that initial conditions could be optimized to make sure that this happens without violating the uncertainty principle \eqref{unc} $\delta\phi_i\delta\dot\phi_i\geq\frac{\Gamma((d+1)/2)H_i^{d-1}}{2\pi^{d-1/2}}$. 

For the initial conditions
\begin{align}
\delta\phi&=\frac{c^\frac{1}{2}\Gamma(\frac{d+1}{2})^\frac{1}{2}H^\frac{d-1}{2}}{2^\frac{3}{4}\pi^\frac{d-1}{4}|V''|_{max}^\frac{1}{4}},\nonumber\\
\delta\dot\phi&=\frac{\Gamma(\frac{d+1}{2})^\frac{1}{2}H^\frac{d-1}{2}|V''|_{max}^\frac{1}{4}}{c^\frac{1}{2}2^\frac{1}{4}\pi^\frac{d-1}{4}},
\end{align}
the uncertainty principle gets saturated and the right hand side of \eqref{40} becomes equal to
\begin{align}
B_1(d)\frac{c^{1/2}}{1-c^2}\frac{V_{max}^\frac{d-1}{4}}{|V''|_{max}^\frac{1}{4}},
\end{align}
where we used the Friedmann equation $(d-1)(d-2)H_i^2/2=V_{max}$. According to \eqref{assum}, the above expression is less than $\Delta\phi$. Therefore, for these initial conditions, $\phi\in[\phi_0,\phi_0+\Delta\phi]$ for every $t\leq c\sqrt{2/|V''|_{max}}$. If we set $t=c\sqrt{2/|V''|_{max}}$, from \eqref{lifetime} we find
\begin{align}
c\sqrt\frac{2}{|V''|_{max}}&\leq-\frac{1}{H}\ln(H)\nonumber\\
&\leq \sqrt\frac{(d-1)(d-2)}{2V_{min}}\ln(\frac{\sqrt\frac{(d-1)(d-2)}{2}}{\sqrt {V_{min}}}),
\end{align}
which can be rearranged into
\begin{align}
\frac{|V''|_{max}}{V_{min}}\geq c^2B_2(d)\ln(\frac{B_3(d)}{\sqrt {V_{min}}})^{-2},
\end{align}
which is our desired result.

Now we use the inequality \eqref{Aunst} that we just proved to obtain a result for quadratic potentials. Suppose the quadratic potential $V(\phi)$ has local maximum $V(\phi_0)=V_0$ and second derivative $-|V''|$ over a field range $[\phi_0,\phi_0+\sqrt\frac{2(1-c)V_0}{|V''|}]$ for some $0\leq c\leq 1$. This field range corresponds to the potential range $[V_{\min},V_0]$ where $V_{\min}=cV_0$. Let $k$ be positive number smaller than $1$. We can weaken the \eqref{Aunst} by multiplying the right hand side of the second inequality by $k$ as
\begin{align}\label{A21}
\Delta\phi< \frac{B_1(d)B_2(d)^\frac{3}{4}V_{max}^\frac{d-1}{4}V_{min}^\frac{3}{4}\ln(\frac{B_3(d)}{\sqrt{V_{min}}})^\frac{1}{2}}{V_{min}B_2(d)-|V''|_{max}\ln(\frac{B_3(d)}{\sqrt{V_{min}}})^2},~~or ~~~\frac{|V''|_{max}}{V_{min}}\geq kB_2(d)\ln(\frac{B_3(d)}{\sqrt{V_{min}}})^{-2}.
\end{align}
If the second inequality gets violated, we get an upper bound on $|V''|$ in terms of $V$. Plugging this upper bound in the first inequality in \eqref{A21} would weaken the above statement to 
\begin{align}
\Delta\phi< \frac{B_1(d)B_2(d)^\frac{3}{4}V_{max}^\frac{d-1}{4}V_{min}^\frac{3}{4}\ln(\frac{B_3(d)}{\sqrt{V_{min}}})^\frac{1}{2}}{(1-k)V_{min}B_2(d)},~~~or~~~\frac{|V''|_{max}}{V_{min}}\geq KB_2(d)\ln(\frac{B_3(d)}{\sqrt{V_{min}}})^{-2}.
\end{align}
By plugging $\Delta\phi=\sqrt\frac{2(1-c)V_0}{|V''|}$ and $V_{\min}=cV_0$ into the above inequalities we find either 
\begin{align}
\frac{|V''|}{V_0} > \frac{2(1-k)^2(1-c)c^\frac{1}{2}B_2(d)^\frac{1}{2}}{B_1(d)^2}V_0^\frac{2-d}{2}\ln(\frac{B_3(d)}{\sqrt{cV_0}})^{-1},~~~or~~~\frac{|V''|_{max}}{V_{0}}\geq kcB_2(d)\ln(\frac{B_3(d)}{\sqrt{cV_0}})^{-2}.
\end{align}
In other words, 
\begin{align}
\frac{|V''|}{V_{0}}\geq\min(kcB_2(d)\ln(\frac{B_3(d)}{\sqrt{cV_0}})^{-2},\frac{2(1-k)^2(1-c)c^\frac{1}{2}B_2(d)^\frac{1}{2}}{B_1(d)^2}V_0^\frac{2-d}{2}\ln(\frac{B_3(d)}{\sqrt{cV_0}})^{-1})
\end{align} 
We can optimize the above inequality by setting $k=1+D(V_0,d)-\sqrt{D(V_0,d)^2+2D(V_0,d)}$ where 
\begin{align}
D(V_0,d)=\frac{c^\frac{1}{2}B_2(d)^\frac{1}{2}B_1(d)^2V_0^\frac{d-2}{2}}{4(1-c)}\ln(\frac{B_3(d)}{\sqrt{cV_0}})^{-1},
\end{align}
so that the two expressions in the $\min(,)$ become equal to each other. This gives
\begin{align}
\frac{|V''|}{V_{0}}\geq (1+D(V_0,d)-\sqrt{D(V_0,d)^2+2D(V_0,d)})cB_2(d)\ln(\frac{B_3(d)}{\sqrt{cV_0}})^{-2}.
\end{align}
Note that the right hand side only depends on $V_0$. This is a potential dependent lower bound on $|V''|/V_0$ for quadratic potentials defined over a potential range $[cV_0,V_0]$ for some number $0\leq c\leq 1$.

\section{Uncertainty Principle}\label{AB}
In this appendix we derive the uncertainty inequality for $\delta\phi\delta\dot\phi$ where $\delta\phi=\sqrt{\expval{(\phi-\phi_0)^2}}$ and $\delta\dot\phi=\sqrt{\expval{\dot\phi^2}}$. Note that since we study the evolution of a Hubble patch, the field values that we work with are not the local field values $\phi(x)$, instead they are averaged over a $(d-1)$-ball of radius $1/H$. 

If we quantize a scalar field in a generic background, using a foliation $\Sigma(t)$ such that $\Sigma$'s are Cauchy surfaces, for every $x_{d-1}\in\Sigma(t)$ and every function $f$ on $\Sigma(t)$, the commutation relations would look like,
\begin{align}\label{quant}
\int_{\Sigma(t)} f(x'_{d-1})[\hat\phi(x_{d-1}),\partial_\mu\hat\phi(x'_{d-1})]da^\mu_\Sigma(x'_{d-1})=if(x_{d-1}),
\end{align}
where $a^\mu$ is the area vector with respect to the background metric. Suppose the metric take the form  
\begin{align}
ds^2=dt^2-g_\Sigma(t)dx_{d-1}^2.
\end{align}
The equation \eqref{quant} would take the form
\begin{align}
\int_{\Sigma(t)} \sqrt{g_{\Sigma}}f(x'_{d-1})[\hat\phi(x_{d-1}),\partial_t\hat\phi(x'_{d-1})]dx'_{d-1}=if(x_{d-1}),
\end{align}
which can be written as
\begin{align}\label{com1}
[\phi(x),\dot\phi(x')]=i\delta_{\mu_\Sigma}(x-x'),
\end{align}
where $\delta_{\mu_\Sigma}$ is the Dirac delta distribution on $\Sigma$ with respect to the measure $\mu_\Sigma$ induced by $g_\Sigma$. If we define $\bar\phi$, and $\bar{\dot\phi}$ to be the average of $\phi$ and $\dot\phi$ respectively over $M\subset\Sigma$ with respect to $\mu_\Sigma$, integrating \eqref{com1} over $\{(x,x')\in M\times M\}$ leads to
\begin{align}
[\bar\phi,\bar{\dot\phi}]=\frac{i}{\mu_\Sigma(M)}.
\end{align}
If we take $M$ to be a $(d-1)$-ball of Hubble radius $1/H$ in a spatially flat FRW background, we find
\begin{align}
[\bar\phi,\bar{\dot\phi}]=\frac{i}{\frac{\pi^{d-1/2}}{\Gamma((d+1)/2)}(\frac{1}{H})^{d-1}},
\end{align}
 which would result in the uncertainty principle
\begin{align}\label{unc}
\delta\phi_i\delta\dot\phi_i\geq\frac{\Gamma((d+1)/2)H^{d-1}}{2\pi^{d-1/2}}.
\end{align}
\medskip 
\bibliographystyle{unsrt}
\bibliography{References.bib}
\end{document}